\documentclass[man,12pt,letterpaper,hidelinks,floatsintext]{apa7} 

\usepackage[american]{babel}
\usepackage{csquotes} 
\usepackage{url} 
\usepackage{doi} 
\usepackage{ragged2e} 
\usepackage{amsmath,amsmath,amsthm,amssymb,dsfont}
\usepackage{booktabs}
\usepackage{tabularx}
\usepackage{tabulary}
\usepackage{multirow}
\newcolumntype{s}{>{\setbox0=\hbox\bgroup}c<{\egroup}@{}} 
\usepackage{caption}
\usepackage{subcaption}
\usepackage{etoolbox} 
\usepackage{setspace} 
\AtBeginEnvironment{table}{\singlespacing}
\AtEndEnvironment{table}{\doublespacing} 

\usepackage{credits}

\credit{JS}  {1,1,1,0,1,1,1,0,0,0,1,1,1,1}
\credit{KG}  {0,1,1,0,1,1,0,0,1,0,1,1,1,1}
\credit{AR}  {0,1,1,0,1,1,0,0,1,0,1,1,1,1}
\credit{BZG} {1,1,0,0,0,1,0,0,0,0,1,0,1,0}
\credit{MP}  {1,0,0,1,0,0,0,0,0,1,0,0,0,1}

\usepackage[style=apa, backend=biber]{biblatex}
\title{Evaluating tree-based imputation methods as an alternative to MICE PMM for drawing inference in empirical studies
}
\shorttitle{Evaluating tree-based imputation methods}

\authorsnames[1,1,1,1,1]{Jakob Schwerter*, Ketevan Gurtskaia, Andrés Romero, Birgit Zeyer-Gliozzo,  Markus Pauly}
\authorsaffiliations{{TU Dortmund University}}
\abstract{\justifying
Dealing with missing data is an important problem in statistical analysis that is often addressed with imputation procedures. The performance and validity of such methods are of great importance for their application in empirical studies. While the prevailing method of Multiple Imputation by Chained Equations (MICE) with Predictive Mean Matching (PMM) is considered standard in the social science literature, the increase in complex datasets may require more advanced approaches based on machine learning. In particular, tree-based imputation methods have emerged as very competitive approaches. 
However, the performance and validity are not completely understood, particularly compared to the standard MICE PMM. This is especially true for inference in linear models. In this study, we investigate the impact of various imputation methods on coefficient estimation, Type I error, and power, to gain insights that can help empirical researchers deal with missingness more effectively. We explore MICE PMM alongside different tree-based methods, such as MICE with Random Forest (RF), Chained Random Forests with and without PMM (missRanger), and Extreme Gradient Boosting (MIXGBoost), conducting a realistic simulation study using the German National Educational Panel Study (NEPS) as the original data source.
Our results reveal that Random Forest-based imputations, especially MICE RF and missRanger with PMM, consistently perform better in most scenarios. Standard MICE PMM shows partially increased bias and overly conservative test decisions, particularly with non-true zero coefficients. Our results thus underscore the potential advantages of tree-based imputation methods, albeit with a caveat that all methods perform worse with an increased missingness, particularly missRanger.
}

\begin{document}


\maketitle

\setcounter{page}{1}
\justifying
\noindent
Missing data is a persistent challenge in social science research, with both surveys and administrative data often containing incomplete information \parencite{Rubin1987}. This problem is particularly pronounced in longitudinal studies, which often suffer from survey non-response across waves. If left unaddressed, missing data can compromise the validity of statistical results like biased estimates, and inflated Type I or II errors \parencite{Collins2001}. A simple, albeit suboptimal, strategy is listwise-deletion or single imputation \parencite{AkandeLR2017, Collins2001, Rubin1987}. Fortunately, there is a more effective approach to this problem: multiple imputation \parencite{Collins2001, Rubin1987}. By using statistical models to replace missing observations with plausible values, multiple imputation generates multiple comprehensive datasets. The beauty of this technique is that it allows uncertainty to be estimated while preserving the inherent variability of the data that would otherwise be lost in single imputation. Each of these complete datasets can be subjected to standard statistical analyses. The results of these analyses are then aggregated for a final estimate \parencite{Rubin1987}.

A popular method within the umbrella of multiple imputation techniques is multiple imputation by chained equations (MICE) \parencite{VanBuurenGO2011}. MICE with predictive mean matching (PMM) is thereby one of the most popular strategies \parencite{vanBuuren2018}. MICE PMM is based on a regression method and works by imputing missing values iteratively using regression models, in which each variable with missing data is modeled conditionally on all other variables in the dataset. MICE PMM is a flexible tool that can handle different types of missing data patterns and allows researchers to specify different models for different variables and data scales \parencite{AkandeLR2017, vanBuuren2018}. MICE PMM allows researchers to impute missing values for multiple variables simultaneously while considering their dependency structure. This is important because social science datasets often contain many interrelated variables. MICE PMM is commonly used for imputing missing data in social science research, \parencite[e.g.,][]{GopalakrishnaEA2022, VanGinkelEA2020, ZettlerEA2022, HajovskyCJ2020, HollenbachEA2021, WestermeierG2016, DeFranzaEA2021} and "arguably the most popular way to implement MI" \parencite[p. 2]{CostantiniEA2023} owing to its flexibility, ability to handle different types of missing data patterns, and ability to consider dependencies among variables. MICE PMM can be easily applied (for example) within the statistical software package \texttt{R} \parencite{R2013} using the package \texttt{mice} \parencite{VanBuurenGO2011}, as well as in other statistical software's like \texttt{Mplus}, \texttt{SPSS}, and \texttt{Stata}. 

Despite its popularity, MICE has its limitations. The flexibility inherent in MICE leaves room for potential misspecification, which can lead to biased imputations as there are several methods within the MICE framework to choose from, and most choose PMM \parencite{Hayes2018, vanBuuren2018}. Additionally, MICE PMM struggles with high-dimensional data and becomes inefficient as the number of variables exceeds the number of observations \parencite{Hayes2018}. Tree-based imputation methods are emerging as a potential solution, particularly suitable for handling missing data in contexts characterized by complex interactions and high-dimensional datasets \parencite{Hayes2018}. These techniques can handle mixed data types such as categorical and continuous variables, and provide a nonparametric approach that can capture nonlinear and interaction effects. Another benefit of tree-based methods is their robustness. They partition the data space into regions, reducing the influence of outliers to their respective local regions. In contrast, MICE PMM in its standard implementation, relying on linear regression models, can be sensitive to outliers, potentially leading to biased imputations. Tree-based methods impute missing data based on observed patterns, which is advantageous when the missingness mechanism is not completely random \parencite{HayesM2017}. Lastly, tree-based methods perform well on high-dimensional and complex data, often handling a large number of variables and interactions without overfitting, while MICE PMM may face challenges in such scenarios \parencite{Hayes2018}. Three such promising tree-based method are MICE with Random Forest (RF), Chained Random Forest \parencite[missRanger, ][]{Mayer2019} and Extreme Gradient Boosting \parencite[mixgb, ][]{DengL2023}. Despite, e.g., missRanger's growing use in empirical studies \parencite{Sajeev2022, SchwerterBDM2023metropolitan, Sommer2023, Waggoner2023, ZeyerETunderreview}, a critical question remains unanswered: is statistical inference reliable for data imputed using tree-based methods? This is important because a predecessor of missRanger, the original missForest \parencite{StekhovenB2012} which does not allow for predictive mean matching, has led to inflated Type I errors for specific designs in previous research \parencite{RamosajAP2020}.

Our research aims to fill this gap. Through a realistic simulation study, we conducted a rigorous simulation analysis to investigate the statistical performance of tree-based imputation data, examining differences in coefficient estimation bias, Type I error rate, and power of statistical tests when analyzing the data with Ordinary Least Squares (OLS) regression. The focus of this realistic simulation study was to determine the performance of the tree-based methods compared to the widely used multiple imputation method MICE PMM \parencite{VanBuurenGO2011, CostantiniEA2023}. At the same time, the study aimed to clarify which of the tree-based methods has the better performance. The empirical foundation for our data simulation was carefully constructed based on the National Educational Panel Study (NEPS) data. It provided a rich platform for our realistic simulation study because it is longitudinal and characterized by multiple waves of data for individual respondents. By combining well-established statistical methods with real-world social science data, the study was designed to provide insight into the most effective imputation methods to address the pervasive problem of missing data in social science research with similar panel studies.



The article is organized as follows. First, we briefly describe how we simulated the data. Then, we describe the different multiple imputation methods. Next, we perform the simulation study to illustrate the differences in bias and statistical inference of the different imputation methods. Finally, we provide implications for applied researchers and suggest directions for future research.



\section{Design}\noindent

\subsection{Simulated datasets}\noindent
As we aim for a fair and realistic method comparison, we take up the recent suggestion to follow a mixed approach between simulation and benchmarking, in which the simulation model is motivated from real data, see e.g. \textcite{friedrich2023role,thurow2023simulate}. To this end, our replications are closely based on the National Education Panel Study for the Starting Cohort Adults \parencite[NEPS, SC6: 13.0.0, ][]{neps_network_2022}, which provides data on adult education, professional careers and competence development over the adult life course \parencite{manual2022}. The target population consists of adults living in Germany and born between 1944 and 1986, without regards for nationality, spoken language(s) and labor status \parencite{manual2022}. We used an extract from the NEPS data as a basis to obtain realistic simulated data. In detail, we use the first 5 NEPS waves (numbers 2 to 6, comprising data gathered between 2009 and 2014) and only considered cases that participated in all 5 survey years. We chose the individuals' logarithmic real monthly gross earned income (2020 prices) as the metric outcome variable of our sample analyses and 6 binary, 5 ordinal, 3 nominal and 7 metric variables as predictors. The variables and their summary statistics are presented in Table \ref{tab:datasetfull}. Our final NEPS dataset with complete information consists of 12,410 entries in long format, i.e., 5 waves with 2,482 respondents each.

To simulate the datasets, a mixed approach between simulation and benchmarking was used. Starting from the empirical subset of the NEPS dataset, simulated data were generated while trying to mimic the data distributions and covariance matrix structures from NEPS. Specifically, synthetic data were generated per variable according to 4 groups described in the following subsections and subject to their dependence on other characteristics and/or their variation per wave. For a visual comparison, we plotted all simulated variables for one replication and compared them with the original NEPS distributions in Appendix \ref{sec:A1}.

\begin{table}[htbp!]
\caption{Data description} \label{tab:datasetfull} 
\centering 
\scriptsize
\begin{tabulary}{\textwidth}{lsLsL}

\toprule
\textbf{Variable Name} & \textbf{Type} & \textbf{Description} & \textbf{Observed values} & \textbf{Summary Statistics} \\ 
  \midrule
  \textbf{Metric variables}  \\ \cmidrule{1-1}
  \textit{age} & metric & Age & \{23, …, 67\} & Mean: 46.97, SD: \;\;8.36, OV:  \{23, …, 67\}  \\ 
  \textit{childhh1\_number} & metric & Number of children in household below age of 6 & \{1, 2, 3\} & Mean:\;\; 0.12, SD: \;\;0.39, OV: \{1, 2, 3\} \\ 
\textit{contactattempts} & metric & Number of contact attempts & \{1, …, 200\} & Mean:\;\; 7.43, SD: 10.84, OV: \{1, …, 200\}  \\ 
  \textit{leftright\_2013} & metric & Political position: 0 is left and 10 is right & \{0, …, 10\} & Mean: \;\;5.51, SD: \;\;1.75, OV: \{0, …, 10\} \\ 
  \textit{ln\_real\_inc} &  & Natural logarithm of the monthly salary in EUR & [1.23, 11.11] & Mean: 7.99, SD: 0.68, OV: [1.23, 11.11] \\ 
  \textit{siblings} & metric & Number of siblings & \{0, …, 7\} & Mean: \;\;1.81, SD: \;\;1.54, OV:  \{0, …, 7\} \\ 
    \textit{workinghrs} & metric & Working hours per week & [0, 90] & Mean: 37.06, SD: 11.56, OV: [0, 90] \\ 
  \textit{work\_experience} & metric & Work experience in years & \{0.17, ..., 49.25\} & Mean: 24.01, SD: \;\;8.99, OV: \{0.17, ..., 49.25\} \\ \midrule
  
    \textbf{Binary variables}  \\ \cmidrule{1-1}
  \textit{birthcountry} & binary & Born in Germany & \{Abroad, in Germany\} & Abroad: 5\%; in Germany: 95\% \\ 
  \textit{fixedterm} & binary & Has fixed price contract & \{no, yes\} & No: 96\%; yes: 4\% \\ 
  \textit{ilearn} & binary & Informal learning since last interview & \{0, 1\} & No: 26\%, yes: 74\% \\ 
  \textit{music\_classic} & binary & Listens to classical music & \{0, 1\} & No: 46\%; yes: 54\% \\ 
  \textit{wb} & binary & Further training since the last interview & \{0, 1\} & No: 63\%; yes: 37\% \\ 
  \textit{woman} & binary & Gender: is woman? & \{0, 1\} & No: 52\%, yes: 48\% \\ \midrule
      \textbf{Ordinal variables}  \\ \cmidrule{1-1}
 
  \textit{comp\_size} & ordinal & Company size of the individual's current employee in terms of number of people: less than 5, between 5 and 9, from 10 to 19 and finally between 20 and 99 persons & \{1, 2, 3, 4\} & 1 (Less than 5 persons): 13\%, 2 (5 to under 10 persons): 11\%, 3 (10 to less than 20 persons): 38\%, 4 (20 to under 100 persons): 38\%\\ 
     \textit{education} & nominal & Educational graduation levels grouped into lower secondary, secondary, academic school track ("Gymnasium"), and tertiary degrees & \{Lower Secondary, Secondary, Abitur, University\} & Lower Secondary: 14\%, secondary: 36\%, academic track: 17\%, university: 33\%  \\ 
      \textit{kldb} & ordinal & Kldb (german classification of occupations) requirement for the role in terms of complexity: Low, skilled, complex and highly complex & \{Low, Skilled, Complex, Highly Complex\} & low complexity: 5\%, Skilled: 48\%, complex: 16\%, highly Complex: 31\% \\ 
      \textit{parentsEd} & ordinal & Parents' highest school-leaving qualification: no school-leaving qualification, secondary school diploma obtained in a "Hauptschulabschluss", secondary school diploma from a "Realschulabschluss", High School Diploma (including technical) and other school-leaving qualification. & \{0, 1, 2, 3, 4\} & 0 (No school-leaving qualification): 1\%, 1 (Secondary school diploma from a "Hauptschulabschluss"): 56\%, 2 (secondary school diploma from a "Realschulabschluss"): 20\%, 3 ([technical] high school diploma): 22\%, 4 (Other school-leaving qualification): 0\% \\ 
  \textit{volunteering} & ordinal & Indicates whether the respondent has been involved in Volunteering activities with 3 options: No, Yes, once and More than once & \{0, 1, 2\} & 0 (no): 48\%, 1 (yes, once): 13\%, 2 (yes, several times): 39\%\\ \midrule

      \textbf{Nominal variables}  \\ \cmidrule{1-1}

  \textit{federalstate} & nominal & Federal State in which the individuals are currently living in & \{Schleswig-Holstein, Nordrhein-Westfalen, Niedersachsen, Rheinland-Pfalz, Sachsen, Hessen, Brandenburg, Bayern, Baden-Württemberg, Berlin (Gesamt), Sachsen-Anhalt, Thueringen, Mecklenburg-Vorpommern, Bremen, Hamburg, Saarland\} & Nordrhein-Westfalen: 21.5\%, Bayern: 16.1\%, Baden-Wuerttemberg: 11.8\%, Niedersachsen: 11.0\%, Hessen: 8.8\%, Rheinland-Pfalz: 5.5\%, Sachsen: 4.4\%, Berlin (Gesamt): 3.6\%, Sachsen-Anhalt: 3.3\%, Thüringen: 3.1\%, Brandenburg: 3.1\%, Schleswig-Holstein: 3.0\%,  Hamburg: 1.5\%, Mecklenburg-Vorpommern: 1.4\%, Saarland: 1.2\%, Bremen: 0.7\% \\ 
    \textit{maritalstatus} & nominal & Marital status & \{single, married, divorced, widowed\} & Married: 73\%, single: 19\%, divorced: 6\%, widowed: 1\% \\ 
  \textit{sector} & nominal & Company sector of the individual's current employee & \{ 1, 2, 3, 4\} & 3 (public and personal services): 39\%, 2 (economic services): 30\%, 1 (manufacturing): 30\%, 4 (other industries): 1\% \\  
   \bottomrule
\end{tabulary} 
\begin{minipage}{\textwidth}
{\scriptsize{  \textit{Note:} SD: Standard deviation; OV: Observed values' range. The summary statistics column shows the mean, SD, and OV for metric variables. For the others, it shows how often each observed value occurred. The ordinal and nominal variables are converted to binary variables in the regression, but not for imputation. The German secondary school system follows a track system with three main tracks from the fifth grade onward: \textit{Hauptschule}, \textit{Realschule}, and \textit{Gymnasium} with ascending order, i.e. the students with the greatest academic success go to the Gymnasium.}}
\end{minipage}
\end{table}

\subsubsection{Constant independent features}\noindent
We generated eleven variables which are partly binary, ordinal, and nominal by using random sampling with replacement as available in the sample function from the sampling package \parencite{TilleM2021} in \texttt{R}. The possible variables and their corresponding weights (parameter \textit{p} in \texttt{R}) were set to the probabilities as in the original NEPS dataset distributions. We modeled the following variables: country of birth \textit{birthcountry}, size of current company/employer \textit{comp\_size}, level of education \textit{education}, current federal state of residence \textit{federalstate}, political attitude from 'left' to 'right' \textit{leftright\_2013}, in which sector the current occupation is in \textit{sector}, education level of the parents \textit{parentsED}, the number of siblings \textit{siblings}, and whether the person is volunteering \textit{volunteering}.

Additionally, following the distributions observed in the original NEPS dataset for features \textit{woman} (gender being 1 if female and 0 if male) and \textit{music\_classic} (being 1 if the person listens to classical music and 0 if not), we generated two binary variables following the same approach: for \textit{n} respondents, each person's gender and their response regarding whether they listen to classical music were drawn and modeled as binomial random variables with probabilities of success \textit{p} according to the observations in NEPS \parencite[p.~51]{Rotondi2022}. As expected, the simulations have a good fit to the marginal distributions, see Figure \ref{dplotsA} in Appendix~\ref{sec:A1}.

\subsubsection{Time-varying independent features}\noindent
Next, we used different functions to generate six variables which can change over time. 
Starting with the variable \textit{age}, we observed an approximately truncated normally distributed behaviour in our original NEPS dataset (see the last row of Figure (b) in Appendix~\ref{sec:A1}). We therefore modeled the variable \textit{age} at the first wave as a sum of normal random variables \parencite{pickering2017describing} with truncations to mimic the original distribution more closely  \parencite{robert1995simulation} and then added 1 year for each of the subsequent waves. 
To mimic the the number of contact attempts the individuals had before answering the survey, we used an exponential distribution \parencite{kundu2018univariate} for each wave which shows a good fit 
(see the first row of Figure \ref{dplotsB} in Appendix~\ref{sec:A1}). Thereafter, two variables were generated as  random binary variables with different probabilities of success for each wave using the NEPS variables about individuals' further education \textit{wb} and whether they have informal learning since the last interview \textit{ilearn}. Individual's marital status \textit{maritalstatus} was first modeled as in the previous case by sampling random distributions with weighted probabilities for each value, but then a variance was added for people who might change their marital status, for example from single to married or from married to divorced. This variance was modeled as a random binary distribution with a 5\% probability of moving from one status to the next (note that it was modeled as an ordered nominal variable). %

Similarly, whether individuals have a fixed-term contract was modeled as a random binary variable for the first wave and then we added another random variable with a 30\% probability of success in moving from a fixed-term contract to an open-ended contract to get a close representation of the true variable \textit{fixedterm}. The bar plots from Figure \ref{dplotsB} in Appendix~\ref{sec:A1} again show a good fit for all of these nominal variables to the original marginal distributions.

\subsubsection{Constant endogenous features}\noindent
The empirical distributions observed for the 2 features \textit{kldb} and \textit{workinghrs} did not fit well to any of the classical distributions mentioned before (e.g. binomial, normal, exponential, etc.). Thus, we simulated the data in a different way, using Decision Trees for classification. These relayed on previously generated variables which had correlations to the ones being generated in order to get the closest possible data distribution. In particular, we generated the variable \textit{kldb} using the variables \textit{age, state, education, school, parentsED, company size} and an intercept. Additionally, we generated individuals' working hours (\textit{workinghrs}) on \textit{age, country of birth, female,x comp\_size, federalstate, kldb}, \textit{work\_experience\_years, industry} and \textit{parentsEd}. 

These more complex marginal simulation models do not capture all distributional patterns of the original variables (particularly for the peaks in \textit{workinghrs}) but the fit is nevertheless acceptable regarding the overall tendencies and ranges, see Figure \ref{dplotsC} of Appendix~\ref{sec:A1} for details. 

\subsubsection{Time-varying endogenous features}\noindent
To generate the first of two variables which depend on other variables and change over time, we focused on the individuals' work experience in years \textit{work\_experience}. First, we used a decision tree (dependent on the variables \textit{age, education, marital status, female, federalstate and kldb}) and then added a random uniform variable between 0 and 1 to the previous wave to mimic the evolution on a yearly basis with some randomness for the cases in which the individual was laid off (to preserve the realistic scenario).

Additionally, to generate the number of children in the household the individuals live in, we modeled  the first wave with a Random Forest of 25 trees each and also 3 features to build them, and then a random binary variable was added to mimic the probability of a child becoming older than 6 years (which was set to 20\%). 
The fit of these two complex marginal simulation models is good, see Figure \ref{dplotsD} of Appendix~\ref{sec:A1}.

\subsubsection{Outcome variable}\noindent
The outcome variable mimics individuals' log-transformed income using the natural logarithm. We used a fixed linear model with 16 variables, an intercept and a normally distributed random variable with mean 0 and variance 0.5. The variables chosen were \textit{wave}, \textit{contactattempts}, \textit{workinghrs}, \textit{kldb}, \textit{fixedterm}, \textit{maritalstatus}, \textit{education}, \textit{female}, \textit{age}, \textit{federalstate}, \textit{wb}, \textit{comp\_size}, \textit{sector}, \textit{schoolparents}, \textit{ilearn} and \textit{siblings}. The selection of variables is motivated by a linear regression using the complete NEPS data for which ordinal and nominal variables were transformed to binary variables. That is, variables whose coefficient was estimated to have a statistically significant effect in our complete data were used to generate our outcome variable with their respective estimated coefficient (see Figure \ref{dplotsE} of Appendix~\ref{sec:A1}). The equation is as follows:
{\scriptsize
\begin{equation} \label{eq1} 
\begin{aligned} 
  \textit{ln\_real\_inc} &=  7.1481521 +  0.015  \cdot \textit{wave3} + 0.028 \cdot \textit{wave4} + 0.037 \cdot \textit{wave5} + 0.048 \cdot \textit{wave6} \\
                        &+ 0.001 \cdot \textit{attempts}  +  0.031 \cdot \textit{workinghrs}  - 0.266 \cdot \textit{kldbLow} + 0.155 \cdot \textit{kldbCmplx} + 0.237 \cdot \textit{kldbHighCmplx} \\
                        &- 0.166 \cdot \textit{fixedtermyes}   - 0.058 \cdot \textit{single}   - 0.067 \cdot \textit{LowerSecond}   + 0.067 \cdot \textit{Abitur}  +  0.204 \cdot \textit{University} \\ 
                        &- 0.172 \cdot \textit{woman2}  -  0.006 \cdot \textit{age}  - 0.058 \cdot \textit{FedStatHH}  - 0.033 \cdot  \textit{FedStatBY} - 0.107 \cdot  \textit{FedStatBE} \\
                        &- 0.237 \cdot \textit{FedStatBB} +  0.301 \cdot \textit{FedStatMV} - 0.276 \cdot \textit{FedStatSN} - 0.2000 \cdot \textit{FedStatST} -  0.268 \cdot \textit{FedStatTH} \\
                        &+ 0.050 \cdot \textit{wb2} - 0.185 \cdot \textit{compsize1} - 0.077 \cdot \textit{compsize2} +  0.168 \cdot \textit{compsize4} + 0.060 \cdot \textit{sector1} \\
                        &- 0.169 \cdot \textit{sector4} +  0.059 \cdot \textit{parentsEd3} +  0.057 \cdot \textit{ilearn2}  - 0.014 \cdot \textit{siblings} + \mathcal{N}(0.5, 0.5)
\end{aligned}
\end{equation}
}

\subsubsection{Simulation Evaluation}\noindent
In order to put our simulation results into context for an easier comparison to future simulations, we evaluated the simulated datasets by how good they mimic the subset of NEPS with respect to the correlation structure. Therefore, we calculated the distance between the correlation matrix of the simulated datasets and the original NEPS subset using three common distance measures, namely: Frobenius Distance, Mean Absolute Error and Root Mean Squared Error. For each simulated dataset, the correlation matrix was calculated following the Spearman, Kendall and Pearson approaches 
and the results are shown in Table \ref{tab:variances}. The numbers are quite small and indicate that our simulated dataset is also close to the original dataset with respect to different correlation measures.


\begin{table}[hbpt!]
\caption{Distances between the covariance matrices for the NEPS dataset and the simulated dataset} \label{tab:variances} \vspace{-2em}
\begin{center}
\footnotesize
\begin{tabular}{lrrrrrrrrr}
	\toprule
& \multicolumn{3}{c}{Spearman} & \multicolumn{3}{c}{Kendall} & \multicolumn{3}{c}{Pearson} \\ \cmidrule(lr){2-4} \cmidrule(lr){5-7} \cmidrule(lr){8-10}
	&Frobenius & MAE & RMSE&Frobenius & MAE & RMSE&Frobenius & MAE & RMSE \\ 
	\midrule
	Min.    & 2.027   & 0.0634   & 0.0921   & 1.763 & 0.0549   & 0.0802 & 2.038   & 0.0635   & 0.0927  \\
	1st Q.  & 2.121   & 0.0669   & 0.0964 & 1.843   & 0.0580   & 0.0838 & 2.121   & 0.0664   & 0.0964 \\
	Median  & 2.146   & 0.0679   & 0.0976 & 1.865   & 0.0588   & 0.0848 & 2.149   & 0.0672   & 0.0977 \\
	Mean    & 2.146   & 0.0678   & 0.0976 & 1.864   & 0.0587   & 0.0847 & 2.148   & 0.0673   & 0.0976 \\
	3rd Q.  & 2.171   & 0.0687   & 0.0987 & 1.885   & 0.0595   & 0.0857 & 2.172   & 0.0681   & 0.0987  \\
	Max.    & 2.265   & 0.0717   & 0.1030 & 1.968   & 0.0620   & 0.0895 & 2.261   & 0.0713   & 0.1028  \\
\bottomrule
\end{tabular}
\end{center}
\end{table}

\subsection{Missingness mechanism}\noindent
To simulate real-world scenarios of data incompleteness, we firstly generate a comprehensive dataset as explained in the previous paragraphs, aiming to mimic real-world scenarios as in NEPS. To introduce missingness, we artificially employ three distinct missing rates -10\%, 30\% and 50\%- allowing us to explore the consequences of varying degrees of data incompleteness. These missing data points were constructed using the Missing At Random (MAR) framework. To describe this, consider a dataset with $n$ observations measured on $p$ variables given by 
$Y= (y_{i,j})_{i,j}$. It consists of observed and missing parts $Y=(Y_{obs}, Y_{mis}$) as described by \textcite{Littlerubin}, MAR refers to a scenario in which the probability of a data observation being missing depends only on the observed data ($Y_{obs}$), but not on the missing parts of the data ($Y_{mis}$): 
\begin{align*}
    P(R|Y_{obs}, Y_{mis}, \Phi) = P(R| Y_{obs}, \Phi),
\end{align*} 
where $\Phi$ is the parameter describing the probability of missingness. Assuming MAR makes the unknown missingness parameters negligible in the imputation process. 
To generate MAR missingness in our simulated data, we used the same procedure as described in \textcite{Thurow2021}. It is an iterative method that considers two variables: The first variable introduces missingness completely at random (MCAR) which is then used to generate observations MAR for the second variable by modelling the missing probability dependent of the observed values of the first variable. For the concrete form of the MAR generation model, we refer to part \ref{sec:A2} of the Appendix.

\section{Imputation}\noindent
\subsection{MICE}\noindent
Multiple Imputation by Chained Equations (MICE) is a sophisticated method for imputing missing data. Implemented in the mice package and rooted in the principle of fully conditional specification \parencite{VanBuurenGO2011}, MICE works by decomposing a multidimensional imputation problem into multiple one-dimensional problems. Specifically, the data are decomposed into a target variable and its corresponding predictors. Rather than imputing all missing values simultaneously, MICE systematically imputes one variable at a time, conditional on the observed and imputed values of the other variables.

The mathematical mechanics underlying MICE involves drawing samples from distributions of the form $P(Y_j| Y_{-j},\theta_j)$, where $Y_{-j}$ represents all variables except the $j$-th variable. The posterior distribution of $\theta$ \parencite[the parameter of interest that fully specifies the multivariate distribution underlying $Y$, see][]{VanBuurenGO2011} is derived from these conditional distributions. In each iteration of the MICE algorithm, the imputation process is refined using subsequent draws of the Gibbs sampler: 
\begin{align*}
&\theta^{1(t)} \sim P(\theta^1|Y_1^{obs}, Y_2^{(t-1)}, ..., Y_k^{(t-1)})\\
&Y_1^{(t)} \sim P(Y_1|Y_1^{obs}, Y_2^{(t-1)}, ... Y_k^{(t-1)}, \theta^{1(t)}) \\
&\hspace{1cm} \vdots \\
&\theta^{k(t)} \sim P(\theta^k|Y_k^{obs}, Y_1^{(t)}, ..., Y_{k-1}^{(t)})\\
&Y_k^{(t)} \sim P(Y_k|Y_k^{obs}, Y_1^{(t)}, ... Y_k^{(t)}, \theta^{k(t)}).
\end{align*} 
MICE iterates between all available variables as output variables. This chained approach ensures that the imputations are coherent and well informed by the underlying data structure, often leading to more robust statistical inferences when analyzing the imputed data \parencite{vanBuuren2018}. 

\subsubsection{Predictive Mean Matching}\noindent
\label{pmm}
The imputation method specification Predictive Mean Matching (PMM) in the context of MICE serves as a semi-parametric ad hoc imputation method. One of the main advantages of PMM is its ability to preserve nonlinear relationships in the data. PMM's modus operandi is primarily to compute predicted values for each target variable with missing entries according to the specified model.
With PMM, the imputation process works by first estimating the conditional distribution of the missing variable given the observed variables for each individual with missing data. Rather than directly imputing a single value from this distribution, PMM first forms a so called donor pool from the 
the completely observed data. These donors have predictions similar to the predictions for the missing variable. Then PMM randomly draws the imputation value from the donors. The involved matching process helps preserving the distributional characteristics of the observed data while providing a plausible imputed value for the missing data point \parencite{vanBuuren2018}. When selecting donor pools for imputation, smaller donor pools, such as 5, the default for \verb!mice! \parencite{vanBuuren2018}, tend to produce better results than large donor pools \parencite{Kleinke2017}.


In essence, PMM fills in the missing values not with arbitrary or mean-based values, but by selecting observed values from the same variable or distribution that are closest to the predicted values. This methodological approach ensures that not only the original data distribution is preserved, but also the relationships between variables. Therefore, PMM is particularly useful when dealing with non-normally distributed or skewed data with a sufficiently large sample size \parencite{Moris2014,Kleinke2017}. One of the outstanding advantages of PMM is its robustness: it significantly reduces potential biases in the imputation process and is less prone to model misspecification problems \parencite{vanBuuren2018}. Since PMM reflects the structure of the original data, it provides a reliable way to deal with missing data in the context of MICE. Nevertheless, caution is required before choosing PMM as an imputation method: if the data are extremely skewed or if a substantial portion of the data requires imputation, PMM may not be the best choice for multiple imputation \parencite{Kleinke2017}.

\subsubsection{Random Forest}\label{RF} \noindent 
The Random Forest introduced by \textcite{Breiman} is a machine learning algorithm that is particularly valued for its ability to handle complex relationships, intricate interactions, and high-dimensional data. A Random Forest is an ensemble of decision trees, where each tree in this ensemble functions either as a classifier, for categorizing data into distinct classes, or as a regressor, for predicting continuous outcomes \parencite{RFCutler}. This versatility allows Random Forests to effectively handle a wide range of data types, from categorical to metric. The strength of this method lies in its ensemble approach: by combining the predictions of multiple trees, it significantly improves the accuracy and robustness of the model, reducing the risk of overfitting that is common with individual decision trees. This 
ensures a more reliable and nuanced usage of the underlying data patterns. Each tree is created from a random subset of the training data, with the same length as the training data, a technique commonly referred to as bootstrap aggregation or "bagging". To provide additional randomness and diversity, only a random (true) subset of features is considered at each split point of a decision tree. 
In practice, each tree in the Random Forest makes its own prediction when given a data point as input. In classification tasks, individual trees cast their votes to determine the class label, and the final classification results from the mode of these votes. In regression tasks, however, the predictions of the individual trees are averaged to produce a continuous result \parencite{algorithmbreiman}. 

MICE with Random Forest (MICE RF), uses Breiman's Random Forest algorithm for imputation. As defined in \textcite{DOOVE201492}, the algorithm first divides the dataset into two subsets: one with observed values for the target variable and one with missing values, ordered by number of missing values, allowing models with maximum available information. The algorithm initializes imputations by randomly drawing from observed data and iteratively refining them. For each tree in the Random Forest, a model is trained using the observed values, and this model is then used for predictions. These predictions serve as a guide to impute missing values in the dataset. The process is repeated multiple times to form a forest of trees, and the final imputed values are obtained by sampling from the imputed values generated by individual trees. Different to other MICE approaches such as PMM, this approach thus does not require the use of a simple regression model. 
Therefore, MICE RF often leads to more accurate estimates and narrower confidence intervals than parametric MICE methods when there are nonlinear relationships in the data, and Random Forest has shown the potential to be less biased than parametric MICE \parencite{ShahEA2014}.
Additionally, \textcite{AkandeLR2017} and \textcite{Murray2018} have recommended tree-based methods over linear regression methods in MICE, as already Classification and Regression Trees (CART) show reduced bias under both MCAR and MAR even in the presence of a large proportion of missing values compared to regression methods. However, CARTs can overfit by following the pattern of the noise too closely, resulting in a complex model with poor predictive power on new data sets \parencite{BurgetteR2010, ShahEA2014}. Random Forest is an extension of classification and regression trees \parencite{BurgetteR2010, ShahEA2014} that overcomes this problem, does not rely on distributional assumptions and can account for nonlinear relationships and interactions.

\subsection{Chained Random Forest}\noindent
The imputation algorithm in the R package missRanger, based on fast the implementation of Random Forests \parencite[ranger, ][]{RangerWright2017}  is designed for efficient missing value imputation utilizing Chained Random Forests, as introduced by \textcite{Mayer2019}. This provides an alternative, much faster implementation of the original missForest algorithm introduced by \textcite{StekhovenB2012}. Additional to leveraging Random Forests, this approach has the capability to integrate predictive mean matching. Using the computational efficiency of the ranger package by \textcite{RangerWright2017}, missRanger merges the speed and adaptability  of Random Forest imputation with the advantages offered by predictive mean matching \parencite{Mayer2019}. 
This feature serves two purposes: preventing imputation with values not present in the original data to preserve the support of the empirical distribution \parencite[Chapter 3.4]{vanBuuren2018}, and increasing the variance in conditional distributions to a realistic level \parencite{Mayer2019}. This allows for multiple imputation scenarios when missRanger is called repeatedly. As missRanger is a Random Forest method, it is particularly useful for imputing missing values in mixed and high-dimensional data and for capturing nonlinear interactions between variables \parencite[][see also above]{DOOVE201492}.

Finally, the use of missRanger with predictive mean matching offers several advantages over traditional imputation methods in the context of social science research. Its use of Random Forests and predictive mean matching can effectively capture complex and non-linear relationships between variables, making it suitable for datasets where such interactions are prevalent. This is particularly beneficial in the social sciences, where intricate relationships and nuanced dependencies between variables are common. Predictive mean matching in missRanger works exactly as described in section above in the subsection on PMM, with the addition that the prediction of the donor values is done with Random Forests. missRanger also offers the possibility not to use PMM, this is similar to missForest by \textcite{StekhovenB2012}, but much faster thanks to the ranger implementation of Random Forest.
While MICE has been a reliable and widely used imputation method in the social sciences, missRanger represents a promising alternative that can potentially outperform MICE in scenarios where complex relationships and high-dimensional data are prevalent (i.e., the limitations of MICE).
However, it is important to acknowledge that missRanger is a relatively new package and has not been extensively tested compared to MICE for parameter estimation and testing, and thus requires further validation. In addition, missRanger is explicitly designed for Random Forest imputation, which may not be appropriate for all data types and research questions. 

\subsection{Gradient Boosting}\noindent
Extreme Gradient Boosting (XGBoost), a machine learning algorithm belonging to the gradient boosting family \parencite{ChenG2016}, combines multiple regression trees using gradient boosting, incorporating regularization and shrinkage elements.
XGBoost utilizes gradient descent optimization for efficient loss function minimization, incorporating L1 and L2 regularization to prevent overfitting. Tree pruning reduces model complexity, and parallel processing capabilities make it suitable for large datasets. The algorithm is renowned for handling complex datasets, missing values, and providing accurate predictions, evident in Kaggle competitions and real-world applications \parencite{ChenG2016}.

The mixgb package in R employs XGBoost for scalable missing value imputation \parencite{DengL2023}. The package incorporates features such as subsampling with different ratios and the option to choose the type of Predictive Mean Matching (PMM). Prioritizing variables with fewer missing values, mixgb uses bootstrapping and PMM for accurate imputation, particularly for continuous data. The approach is non-iterative, in contrast to MICE, however allows the flexibility to specify the number of iterations within the function. According to \textcite{DengL2023} first, an initial imputation is conducted to obtain a complete and sorted dataset. Afterwards, XGBoost models with subsampling are employed to address the uncertainty of missing values, akin to sampling parameters of a parametric model from their posterior distributions. The subsampling ratio influences the data subset used in each boosting round. For each incomplete variable, an XGBoost model with subsampling is fitted, and predictions are obtained. mixgb leverages XGBoost's predictive capabilities and parallel processing, benefiting from its diagnostic tools for imputation quality assessment \parencite{DengL2023, SuhS2023}. By default, mixgb utilizes imputations without PMM for categorical variables, while employing PMM for numeric variables with a standard donor count set at five.

\subsection{Imputation Method Evaluation}\noindent
We used ten imputations as increasing the number of imputations did not change the results considerably in other simulation studies \parencite{ShahEA2014, Schafer1999}. We replicated the simulation 1200  times, taking the first 1000 that worked. For each design, there were some replications that failed at the very beginning when running the simulation on the cluster (\texttt{LiDO3}) at TU Dortmund University. The most common failure was that the simulated datasets did not have all the possible values in categorical features due to low probabilities to be drawn.

We will compare the different methods with respect to different criteria, analyzing their bias in coefficient estimation as well as the validity of subsequent statistical tests, measured in terms of Type I error rates and power. In addition, we use the Imputation Performance Metric (IPM), which is a single summarizing measure proposed by \textcite{SuhS2023} to simplify the comparison of different imputation methods. Unlike the RMSE used for numerical datasets and the misclassification error for categorical datasets \cite[e.g.,][]{ramosaj2019predicting}, this novel metric method is able to combine mixed datasets of both types into a single metric, making it suitable for comparing the imputation results obtained for our selected subset of variables from NEPS, which includes metric, binary, ordinal, and nominal variables. The metric is calculated as follows:
Continuing with the notation from the previous section, let $Y = (y_{ij})$ be a dataset with $i=1,...,n$ observations and $j=1,...,p$ variables. Let $Y^{T}$ be the true dataset before adding missingness and $Y^{*}$ be the single imputed dataset (after adding $n^{mis}$ missing entries). 
Now define the distance between two numeric columns $j$ of $Y^{T}$ and $Y^{*}$ as
$$\delta \big( Y_{ij}^{T} - Y_{ij}^{*}\big) = \frac{|Y_{ij}^{T} - Y_{ij}^{*}|}{ \max (Y_{j}^{T},Y_{j}^{*}) - \min (Y_{j}^{T},Y_{j}^{*}) }$$
and analogously the distance between two categorical columns $j$ of $Y^{T}$ and $Y^{*}$ as
$$\delta \big( Y_{ij}^{T} - Y_{ij}^{*}\big) = \mathds{1} \big(Y_{ij}^{T} \neq Y_{ij}^{*} \big),$$
where $\mathds{1}(\cdot)$ is 1 if $Y_{ij}^{T} \neq Y_{ij}^{*}$ are different and 0 if they are equal. Then IPM is defined as
$$IPM = \sum_{i=1}^n \sum_{j=1}^p \frac{\delta \big( Y_{ij}^{T} - Y_{ij}^{*}\big)} {n^{mis}},$$
which means that as a measure of distance, the lower the value for IPM, the closer the imputed dataset $Y^*$ is to the true one $Y^T$, and the bigger it is, the more different they are. A perfect imputation would therefore result in an IPM value of 0.

This metric comes with the advantage of aggregating results from variable data types into one single number. 
At the same time this comes with the 
drawback, that the aggregation might not be fair among all of the variable types. For example, the effect of miscalculated metric variables that are not that apart from each other will be lower (the $\delta \big( Y_{ij}^{T} - Y_{ij}^{*}\big)$ closer to 0) than the effect of the gender being mistakenly assigned as 1 when the true value is meant to be 0 (in these cases it will always be $\delta \big( Y_{ij}^{T} - Y_{ij}^{*}\big) = 1$).

\section{Results}\noindent
\subsection{Coefficient estimation}\noindent
Figure \ref{fig:bias} gives a visual representation of the coefficient bias for simulations with a missing rate of $10\%$, $30\%$, and $50\%$ for all non-true zero coefficients (i.e. the corresponding variable has an effect on the outcome) on the left and true zero coefficients on the right (i.e. the corresponding variable has no effect on the outcome). This allows for an easy assessment of how different imputation methods affect coefficient estimates. 

\begin{figure}[htbp!]
    \centering
    \caption{Coefficient Bias under MAR} \label{fig:bias}

    \begin{subfigure}{\textwidth}
        \centering
        \subcaption{Coefficient Bias for 10\% missingness} \label{fig:bias10}
        \includegraphics[width=0.95\textwidth]{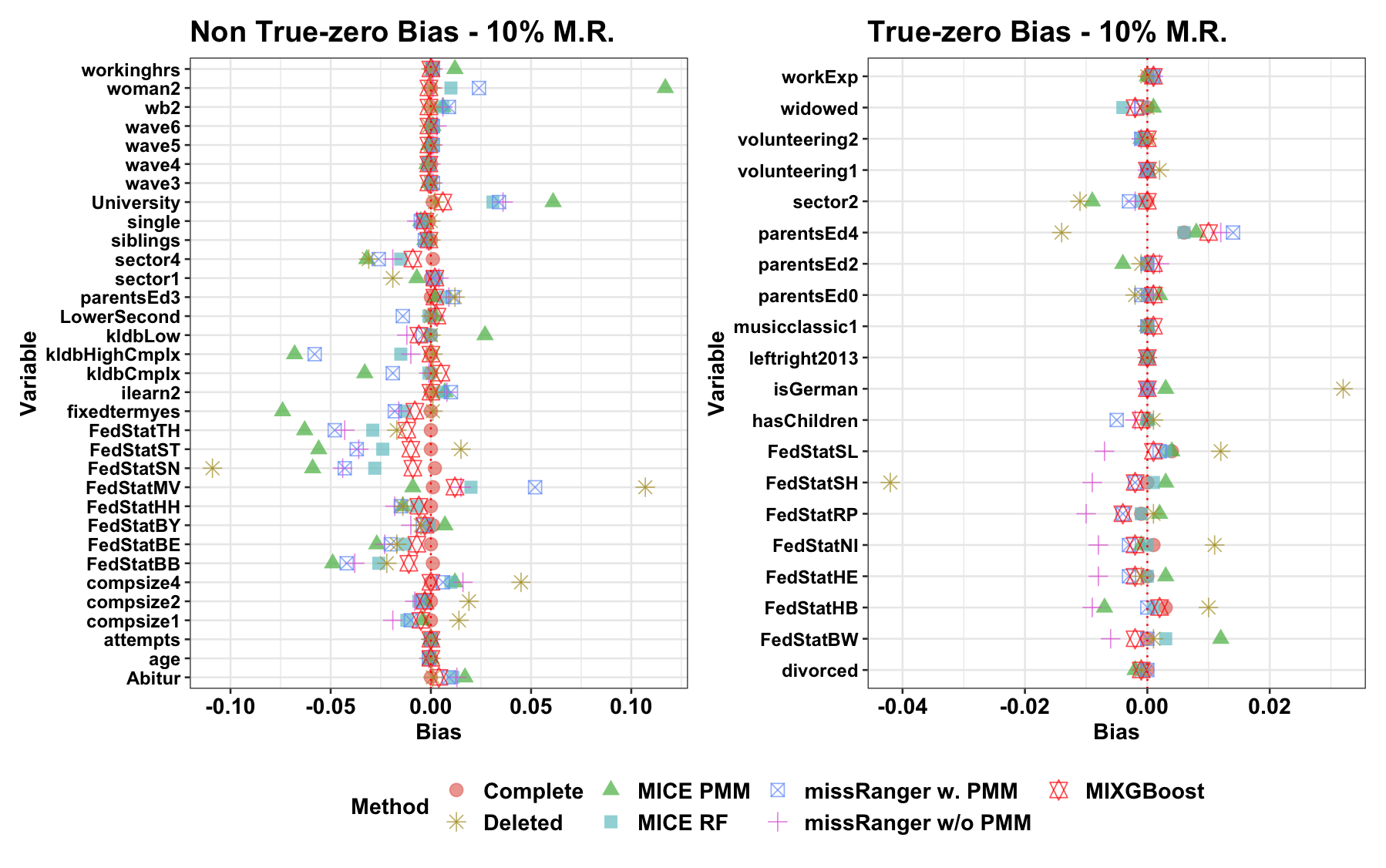}
    \end{subfigure}

    \begin{subfigure}{\textwidth}
        \centering
        \subcaption{Coefficient Bias for 30\% missingness} \label{fig:bias30}
        \includegraphics[width=0.95\textwidth]{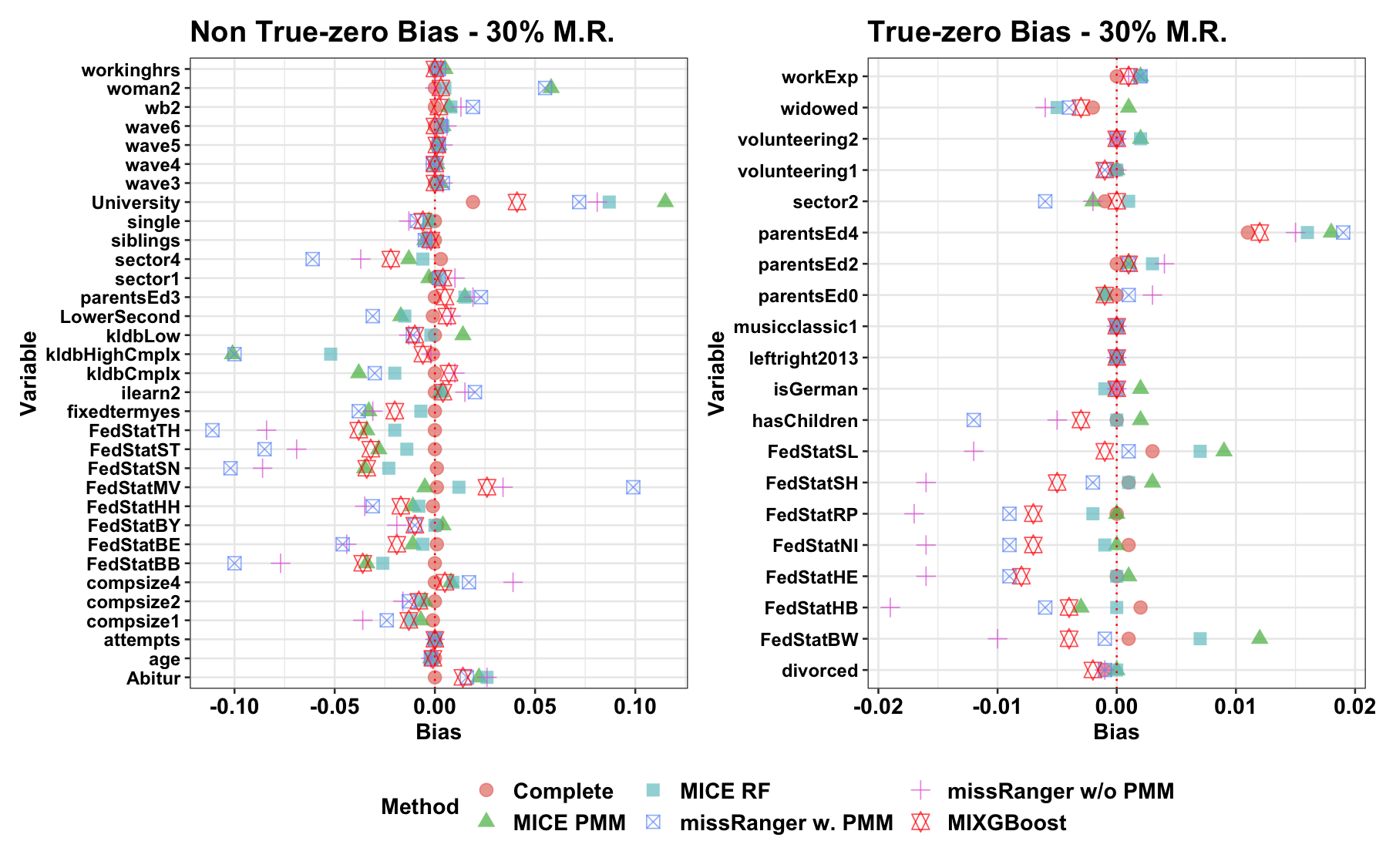}
    \end{subfigure}
\end{figure}

\begin{figure}[htb!]
\ContinuedFloat
    \begin{subfigure}{\textwidth}
        \centering
   \subcaption{Coefficient Bias for 50\% missingness } \label{fig:bias50}
   \includegraphics[width=0.95\textwidth]{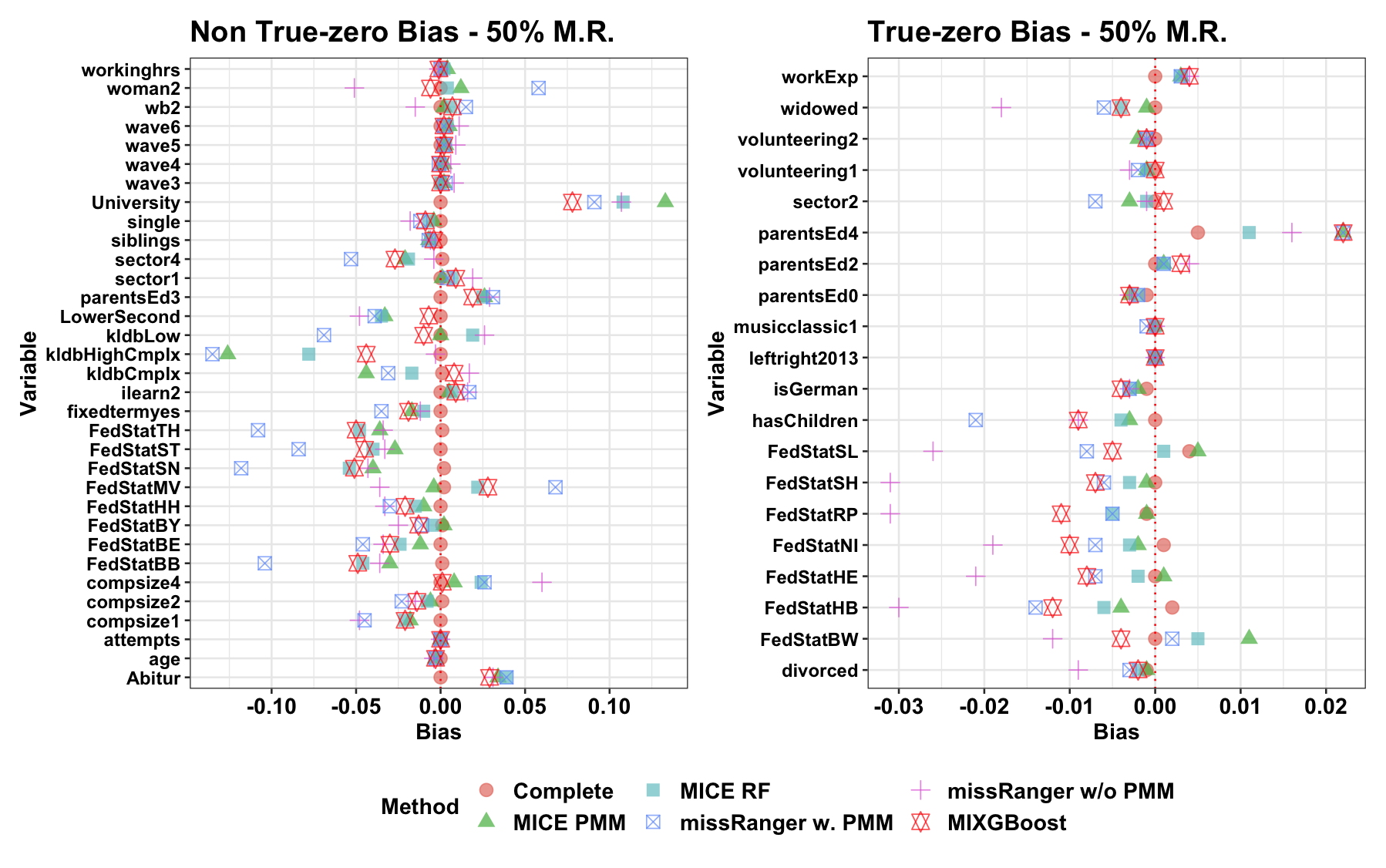}
    \end{subfigure}
\begin{minipage}{\textwidth}
{\scriptsize{\textit{Note:} The figure shows the bias of the coefficient estimates for settings with  (a) $10\%$, (b) $30\%$ and c) 50\% missing rate, with higher values indicating a higher bias. On the left for coefficients not equal to zero (as fixed in Equation (\ref{eq1})), and on the right for coefficients equal to zero. "Complete" refers to estimation before any data are amputated, and "Deleted" refers to regression on listwise deleted datasets. The others colored dots refer to the imputation methods explained in the text. For the 50\% missingness rate, list-wise deletion (Deleted) is not possible as there a not enough observations with full information.}}
\end{minipage}
\end{figure}


Subfigure \ref{fig:bias10} shows that even for a small missingness rate ($10\%$), list-wise deletion (Deleted) has a high bias, especially for true zero coefficients. This can be seen within the graph, as the symbol for list-wise deletion is most often furthest away from the zero line, indicating no bias. Thereafter, MICE PMM  has the highest bias among the imputation methods, closely followed by missRanger without PMM. MICE RF has most often the smallest bias. In general, the biases vary in absolute values between the different coefficient estimates and are not equally distributed.

 Looking at the case with 30\% and 50\% missing rate in Subfigures \ref{fig:bias30} and \ref{fig:bias50}, the bias increases for almost all methods, but only by a very modest amount, while it actually decreases for MICE PMM, particularly for 50\% missingness. While for 30\% missigness it is somewhat mixed, for 50\% missingness, MICE PMM has the lowest bias for most coefficients, and one of the two missRanger implementations has the highest bias. 

To summarize the graphs, we pooled the biases by taking the mean, median, and standard deviation, as presented in Table \ref{allaverage}, over all variables (Panel A), all non true-zero coefficients (Panel B), all true-zero coefficients (Panel C), all coefficients of metric variables (Panel D), and all coefficients of binary variables (Panel E); columns (1) to (3) for 10\% missingness, columns (4) to (6) for 30\% missingness, and columns (7) to (9) for 50\% missingness. Pooling the biases, MICE PMM has the highest average bias of $M_{bias}=0.0326$ and also the highest standard deviation of $SD= 0.1252$ for 10\% missingness (Table \ref{allaverage} Panel A, columns (1) to (3)). This suggests that MICE PMM varies the most in bias values among all the imputation methods.  The tree-based methods missRanger without PMM ($M=0.0179$, $SD=0.0630$), missRanger with PMM ($M=0.0191$, $SD=0.0630$), and MICE RF ($M=0.0149$, $SD=0.0643$) are very close to each other for 10\% missingness. MIXGBoost has the lowest bias and standard deviation ($M=0.0107$, $SD=0.0560$). The low standard deviation values indicate that these methods are relatively consistent and more accurate than MICE PMM.

For 30\% missingness, missRanger with PMM has the highest bias ($M=0.0315$, $SD=0.0693$), which, however, is driven by some estimates of the federal states variables, as one can see by Figure \ref{fig:bias30} and that the median is smaller than for missRanger without PMM. MIXGBoost once again has the smallest estimation bias and smallest standard deviation ($M=0.0159$, $SD=0.0556$), closely followed by MICE RF ($M=0.0161$, $SD=0.0573$). The order of the methods in terms of highest and lowest bias for 50\% missingness is similar to 30\% with a general increase in the bias (except for missRanger without PMM).

\begin{table}[htb!]
\caption{Aggregated Bias of all Methods and Missingness Rates} \label{allaverage} 
\scriptsize
\begin{tabularx}{\textwidth}{Xrrrrrrrrr}
\toprule
& \multicolumn{3}{c}{\textbf{10\% Missingness}} & \multicolumn{3}{c}{\textbf{30\% Missingness}} & \multicolumn{3}{c}{\textbf{50\% Missingness}} \\ 
\cmidrule(lr){2-4} \cmidrule(lr){5-7} \cmidrule(lr){8-10}
  & \multicolumn{1}{c}{Mean} & \multicolumn{1}{c}{Median} & \multicolumn{1}{c}{SD} & \multicolumn{1}{c}{Mean} & \multicolumn{1}{c}{Median} & \multicolumn{1}{c}{SD} & \multicolumn{1}{c}{Mean} & \multicolumn{1}{c}{Median} & \multicolumn{1}{c}{SD} \\
 & \multicolumn{1}{c}{(1)} & \multicolumn{1}{c}{(2)} & \multicolumn{1}{c}{(3)} & \multicolumn{1}{c}{(4)} & \multicolumn{1}{c}{(5)} & \multicolumn{1}{c}{(6)} & \multicolumn{1}{c}{(7)} & \multicolumn{1}{c}{(8)} & \multicolumn{1}{c}{(9)} \\
\midrule
\multicolumn{5}{l}{\textbf{Panel A: Overall Coefficient Bias}} & & & & \\ \cmidrule{1-3}
Complete data & 0.0084 & 0.0000 & 0.0585 & 0.0090 & 0.0000 & 0.0589 & 0.0084 & 0.0000 & 0.0585\\
List-wise deletion & 0.0191 & 0.0010 & 0.0620 & & & & & & \\
MICE PMM & 0.0326 & 0.0045 & 0.1252 & 0.0247 & 0.0045 & 0.0868 & 0.0231 & 0.0040 & 0.0712\\
MICE RF & 0.0149 & 0.0015 & 0.0643 & 0.0161 & 0.0035 & 0.0573 & 0.0222 & 0.0055 & 0.0609\\
missRanger w/o PMM & 0.0179 & 0.0070 & 0.0630 & 0.0276 & 0.0125 & 0.0702 & 0.0255 & 0.0165 & 0.0422 \\
missRanger with PMM & 0.0191 & 0.0040 & 0.0630 & 0.0315 & 0.0090 & 0.0692 & 0.0336 & 0.0115 & 0.0592\\
MIXGBoost & 0.0107 & 0.0020 & 0.0560 & 0.0159 & 0.0045 & 0.0556 & 0.0191 & 0.0080 & 0.0432\\
\midrule
\multicolumn{5}{l}{\textbf{Panel B: True-zero Coefficient Bias}} & & & &  \\ \cmidrule{1-3}
Complete data & 0.0008 & 0.000 & 0.0017& 0.0012 & 0.0000 & 0.0025 & 0.0008 & 0.0000 & 0.0014\\
List-wise deletion & 0.0071 & 0.001 & 0.0114 & & & & & & \\
MICE PMM & 0.0031 & 0.002 & 0.0034 & 0.0030 & 0.0015 & 0.0047 & 0.0034 & 0.0020 & 0.0050\\
MICE RF & 0.0012 & 0.001 & 0.0016 & 0.0024 & 0.0010 & 0.0039 & 0.0029 & 0.0025 & 0.0026\\
missRanger w/o PMM & 0.0040 & 0.002 & 0.0040 & 0.0072 & 0.0045 & 0.0071 & 0.0120 & 0.0090 & 0.0111\\
missRanger with PMM & 0.0020 & 0.001 & 0.0032 & 0.0042 & 0.0015 & 0.0051 & 0.0060 & 0.0040 & 0.0062\\
MIXGBoost & 0.0016 & 0.001 & 0.0022 & 0.0030 & 0.0015 & 0.0033 & 0.0055 & 0.0040 & 0.0054\\
\midrule
\multicolumn{5}{l}{\textbf{Panel C: Non True-zero Coefficient Bias}} & & & &  \\ \cmidrule{1-3}
Complete data & 0.0129 & 0.0000 & 0.0737 & 0.0136 & 0.0000 & 0.0742 & 0.0129 & 0.0000 & 0.0737\\
List-wise deletion & 0.0262 & 0.0010 & 0.0772 & & & & & & \\
MICE PMM & 0.0500 & 0.0105 & 0.1560 & 0.0375 & 0.0095 & 0.1078 & 0.0347 & 0.0090 & 0.0880\\
MICE RF & 0.0230 & 0.0065 & 0.0804 & 0.0241 & 0.0070 & 0.0713 & 0.0336 & 0.0160 & 0.0748\\
missRanger w/o PMM & 0.0261 & 0.0095 & 0.0786 & 0.0396 & 0.0155 & 0.0865 & 0.0334 & 0.0220 & 0.0511\\
missRanger with PMM & 0.0292 & 0.0100 & 0.0780 & 0.0476 & 0.0215 & 0.0834 & 0.0498 & 0.0310 & 0.0699\\
MIXGBoost & 0.0161 & 0.0030 & 0.0704 & 0.0236 & 0.0065 & 0.0692 & 0.0271 & 0.0115 & 0.0530\\
\midrule
\multicolumn{5}{l}{\textbf{Panel D: Coefficient Bias for Metric Variables}} & & & &  \\ \cmidrule{1-3}
Complete data & 0.0000 & 0.0000 & 0.0000 & 0.0000 & 0.0000 & 0.0000 & 0.0000 & 0.0000 & 0.0000\\
List-wise deletion & 0.0001 & 0.0000 & 0.0004 & & & & & & \\
MICE PMM & 0.0021 & 0.0000 & 0.0045 & 0.0023 & 0.0020 & 0.0021 & 0.0029 & 0.0030 & 0.0024\\
MICE RF & 0.0007 & 0.0010 & 0.0008 & 0.0013 & 0.0010 & 0.0015 & 0.0026 & 0.0030 & 0.0021\\
missRanger w/o PMM & 0.0010 & 0.0010 & 0.0008 & 0.0021 & 0.002 & 0.0020 & 0.0034 & 0.0040 & 0.0034\\
missRanger with PMM & 0.0016 & 0.0010 & 0.0018 & 0.0033 & 0.002 & 0.0042 & 0.0049 & 0.0030 & 0.0076\\
MIXGBoost & 0.0004 & 0.0000 & 0.0005 & 0.0010 & 0.001 & 0.0012 & 0.0030 & 0.0030 & 0.0032\\
\midrule
\multicolumn{5}{l}{\textbf{Panel E: Coefficient Bias for Binary Variables}} & & & &  \\ \cmidrule{1-3}
Complete data & 0.0005 & 0.0000 & 0.0012 & 0.0012 & 0.0000 & 0.0032 & 0.0006 & 0.0000 & 0.0010\\
List-wise deletion & 0.0130 & 0.0020 & 0.0234 & & & & & & \\
MICE PMM & 0.0180 & 0.0065 & 0.0260 & 0.0150 & 0.0050 & 0.0239 & 0.0157 & 0.0045 & 0.0274\\
MICE RF & 0.0070 & 0.0030 & 0.0088 & 0.0096 & 0.0050 & 0.0153 & 0.0162 & 0.0065 & 0.0221\\
missRanger w/o PMM & 0.0108 & 0.0080 & 0.0118 & 0.0210 & 0.0140 & 0.0238 & 0.0229 & 0.0185 & 0.0197\\
missRanger with PMM & 0.0122 & 0.0040 & 0.0158 & 0.0262 & 0.0105 & 0.0334 & 0.0304 & 0.0145 & 0.0357\\
MIXGBoost & 0.0035 & 0.0020 & 0.0036 & 0.0097 & 0.0055 & 0.0113 & 0.0153 & 0.0090 & 0.0174\\
\bottomrule
\end{tabularx}
\begin{minipage}{\textwidth}
{\scriptsize{  \textit{Note:} SD: Standard Deviation. For Panel A, we took the mean and median coefficient bias of all variables and replications to have an aggregated parameter to simply Figure \ref{fig:bias}. The standard deviation shows how the mean bias varies between the different variable coefficients. Panel B to E just uses a subset of variables as described in the panel header to show how the parameters differ for true-zero and non true-zero coefficients, as well as for coefficients of metric and binary variables. Again, for  30\% and 50\% missing rates, list-wise deletion (Deleted) is not possible as there a not enough observations with full information.}}
\end{minipage}
\end{table}

For true zero coefficients, as shown in Figure \ref{fig:bias} and Table \ref{allaverage} Panel B, the bias is generally much smaller than for non true-zero coefficients. For the true-zero coefficients, missRanger without PMM has the highest average bias and standard deviation among the imputation methods (10\% miss: $M=0.0040$, $SD=0.0040$; 30\% miss: $M=0.0072$, $SD=0. 0071$; 50\% miss: $M=0.0120$, $SD=0.0111$), and MICE RF (not MIXGBoost) has the lowest (10\% miss: $M=0.0012$, $SD=0.0016$; 30\% miss: $M=0.0024$, $SD=0.0039$; 50\% miss: $M=0.0029$, $SD=0.0026$). The order of methods in terms of bias for non true-zero coefficients is similar to the overall bias. A visual depiction of the summary results are further presented in Figure \ref{fig:devBias} in the appendix, highlighting how the mean and median change per method over the three different missingness rates.

All methods show on average higher biases for binary variables than for metric variables. The binary coefficient biases also have more variability in the estimates (higher standard deviations, Table \ref{allaverage} Panel D and E).

Comparing the biases between Panels B to E, coefficients of binary variables with non-zero coefficients drive the overall biases. Underscoring the point made above that the plots show that the results are driven by some estimates and not that the biases are evenly distributed across all coefficient estimates, one can also see that the median biases are always much smaller than the average. A visual depiction of the summary results are further presented in Figure \ref{fig:devBias}, highlighting that, e.g., the performance does not necessarily decrease with the missingness rate, but that for some MICE RF it improves, and MIXGBoost is the best for 30\% missingness.

\subsection{Test decision}\noindent
Similar to bias, Figure \ref{fig:rejection1} shows for the different missingness rates a per-coefficient overview of $H_0$ ($\beta = 0$) rejection rates across imputation methods, on the right for Type I error rates (the prescribed level is 5\%) and on the left for statistical power. Table \ref{decaveragezero} presents the pooled rejection rate results for 10\% missingness across coefficients in Panel A (Type I) and Panel B (statistical power). The rejection rates measure the frequency with which estimated coefficients are found to be statistically significant and provide critical insight into the performance of each imputation method in maintaining the validity of statistical inference tests. On the one hand, the rejection rates for true zero coefficients indicate the Type I error of the test, and on the other hand, the rejection rates for non true zero coefficients indicate the statistical power of the test. 

\subsubsection{Type I Error Control}\noindent
Subfigure \ref{fig:rejection10} shows the effect of different imputation strategies on the reliability of statistical inference for 10\% missingness. Specifically, the right plot analyzes how imputation effects Type I error level control, providing a key measure of the method's performance in terms of inference statistics. Here, values below (above) the chosen significance threshold of $5\%$ indicate a rather conservative (liberal) decision. Thus, MICE PMM is very conservative while missRanger without PMM is too liberal for almost all settings. missRanger with PMM improves upon that but exhibits some variation around the 5\% threshold and a tendency of being more conservative than liberal. In contrast, MICE RF and MIXGBoost are more accurate with a slight tendency towards being more conservative (MICE RF) or liberal (MIXGBoost), respectively. 

Summarizing the results of all coefficients with 10\% missingness (Table \ref{decaverage}), MICE PMM has the lowest Type I errors (mean rejection rate $M=0.0080$), while missRanger without PMM has the highest Type I errors ($M=0.0640$) and almost always fails to meet the desired threshold of $5\%$ rejection rate, followed by MIXGBoost ($M=0.0520$). MICE RF stays below $5\%$ for all coefficients ($M=0.0275$), and missRanger with PMM sometimes exceeds the threshold, but still has an average rejection rate below the 5\% threshold ($M=0.0383$). The list-wise deletion has a higher Type I error ($M=0.0683$) than missRanger without PMM (Table \ref{decaveragezero} Panel B, columns (1) to (3)). 

Similarly, for the 30\% in Subfigure \ref{fig:rejection30} and for 50\% Subfigure \ref{fig:rejection50}, missRanger without PMM is again the worst performer, never having a Type I error less than or equal to the required $5\%$ (30\%: $M = 0.1032$; 50\% $M=0.2096$, see Table \ref{decaveragezero}). MICE PMM is again more conservative, with an average rejection rate of $0.0180$ for 30\% missingness and $0.0043$ for 50\% missingness. missRanger with PMM has a Type I error slightly below the desired $5\%$ threshold for 30\% missingness ($M=0.0464$) and slightly above for 50\% missingness ($M=0.0639$), but has a Type I error for most coefficients (30\%: median $=0.0355$; 50\% median $=0.0210$) below .05, albeit with relatively high variability (30\%: $SD=0.0546$; 50\%: $SD=0.1197$). MICE RF remains the most reliable strategy for the 30\% and 50\% missingness rate (30\%: $M = 0.0497$; 50\% $M=0.0386$), exceeding the rejection rate of $5\%$ only once for the 50\% missingness. While MIXGBoost lies only slightly above the .05 treshold for 30\% missingess ($M = 0.0617$), it is clearly above it for 50\% missingness ($M = 0.1067$). A visual depiction of the summary results are further presented in the left plots of Figure \ref{fig:devTD} in the Appendix, highlighting that missRanger without PMM suffers the most (wrt Type I error control) under higher missingness and that MIXGBoost is only reliable for low or medium missingness rates. The others are mostly stable, though MICE PMM becomes even more conservative with an increase in the missingness rate.

We note that the slight variation around 5\% in case of the analysis for the complete data set (before inserting missingness) can be explained by the non exactness of the statistical test and Monte Carlo error.

\subsubsection{Statistical Power}\noindent
Regarding the tests' power, the left plot
 of Subfigure \ref{fig:rejection10}, it is evident that for non true-zero coefficients, the variation is even larger. In particular, almost no imputation method reaches always a power of the desired power of the tests applied to the data set before missings were artificially inserted (represented by the red dot,``Complet'').

For 10\% missingness, as expected from the conservative Type I error control, MICE PMM tends not to reject $H_0$ (average rejection rate $M=0.5722$, see Table \ref{decaverage}, Panel B), MICE RF is less conservative than MICE PMM but also shows too low statistical power for some scenarios (average rejection rate $M=0.7771$). On the other hand, both missRanger variants and MIXGBoost have the highest power (0.8759 without PMM and 0.8554 with PMM, 0.8830 for MIXGBoost), which is very close to the power of the complete data before the amputation (M = 0.8937). It is worth noting that for both missRanger variants and MIXGboost, more than half of the coefficients have a rejection rate higher than $99\%$ (median $0.9990$ and $0.9925$ for missRanger without and with PMM, respectively; 0.9980 for MIXGBoost; 1.0000 for complete). Additionally, for MICE RF, almost half of the coefficients have a rejection rate of almost $90\%$  (median $0.8985$), while MICE PMM still lacks power when looking at the median rejection rates (median = 0.6255; Table \ref{decaveragezero} Panel B). The list-wise deletion has a slightly lower power (average rejection rate $M=0.5598$) than MICE PMM (Table \ref{decaveragezero} Panel B, columns (1) to (3)).

MICE PMM is once again the most conservative while rejecting $H_0$ for $30\%$ missingness ($M =0.5545$, see Table \ref{decaveragezero} Panel B). MICE RF also loses some power with increasing missingness rate ($M=0.6981$). missRanger with PMM loses the most power and has only slightly higher power than MICE RF ($M=0.7159$). missRanger without PMM and MIXGBoost still have the highest power ($M=0.8300$ and $M=0.8425$ respectively). Looking at the median, MICE PMM still has the lowest statistical power. It is the only method rejecting most of the non-zero coefficients less than $80\%$ of the time (median $=0.6520$), whereas for the other methods at least half of the coefficients have a rejection rate around $85\%$ for MICE RF (median $=0.8470$) and missRanger with PMM (median $=0.8785$), and higher than $95\%$ for missRanger without PMM and MIXGBoost (median$=0.9740$ for both). 

\begin{figure}[htbp]
    \caption{$H_0$ rejection rates  under MAR } \label{fig:rejection1}
    \begin{subfigure}{\textwidth}
        \centering
    \subcaption{Rejection rates for $10\%$ missingness} \label{fig:rejection10}
\includegraphics[width=0.95\textwidth]{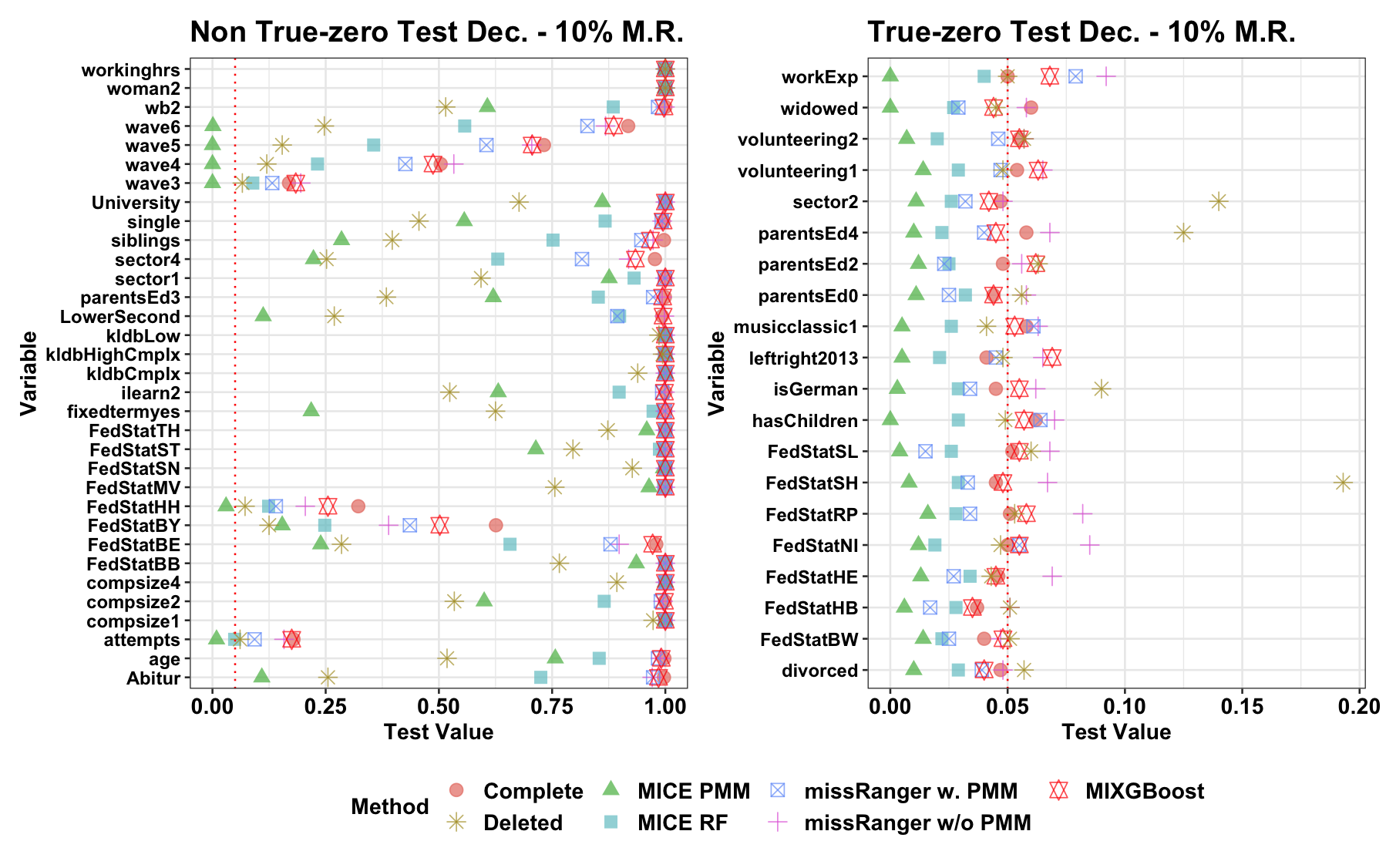}
\end{subfigure}
    \begin{subfigure}{\textwidth}
        \centering
    \subcaption{Rejection rates for $30\%$ missingness} \label{fig:rejection30}
 \includegraphics[width=0.95\textwidth]{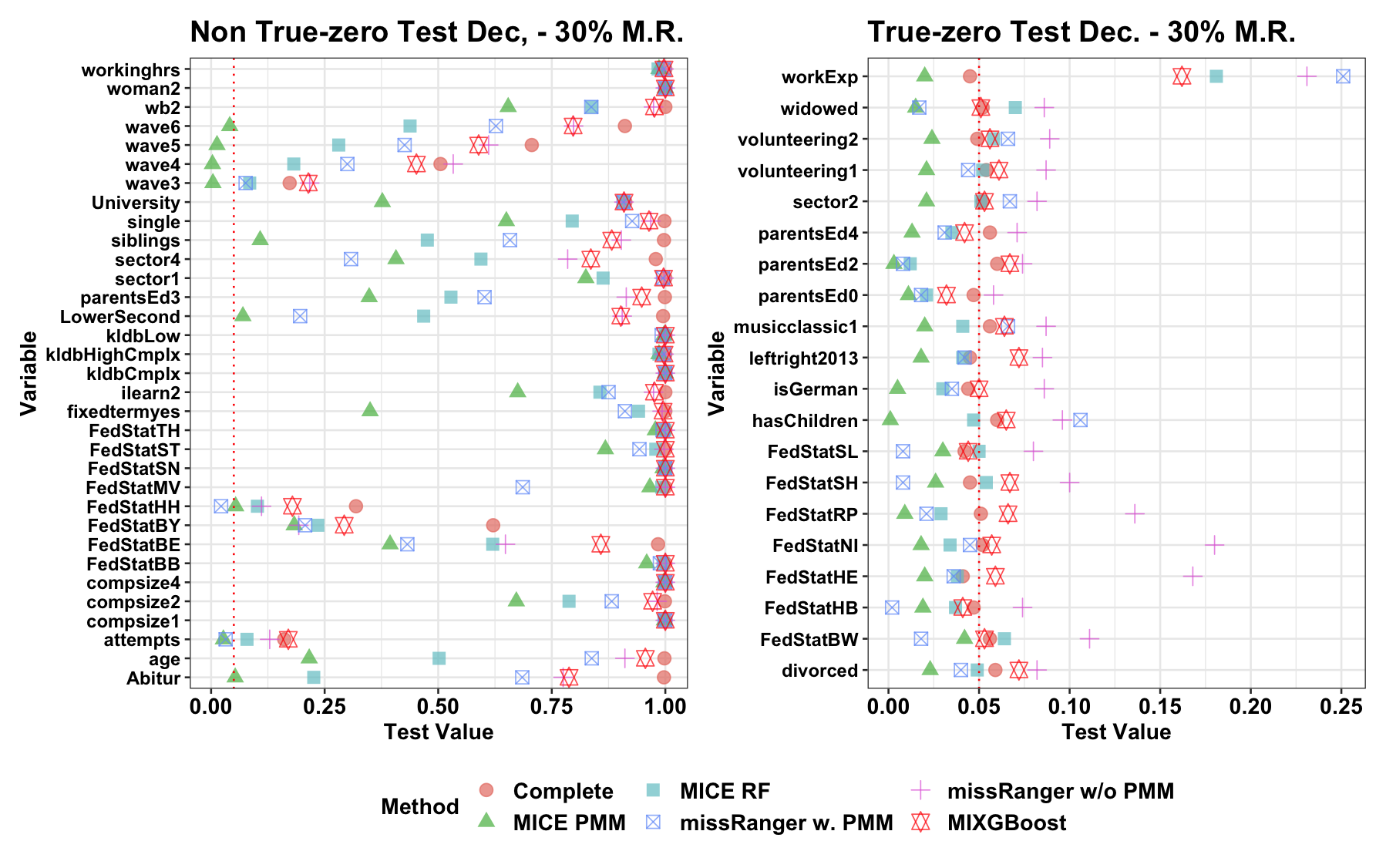}
 \end{subfigure}
\end{figure}

\begin{figure}[htb]
\ContinuedFloat
    \begin{subfigure}{\textwidth}
        \centering
    \subcaption{Rejection rates for $50\%$ missingness} \label{fig:rejection50}
 \includegraphics[width=0.95\textwidth]{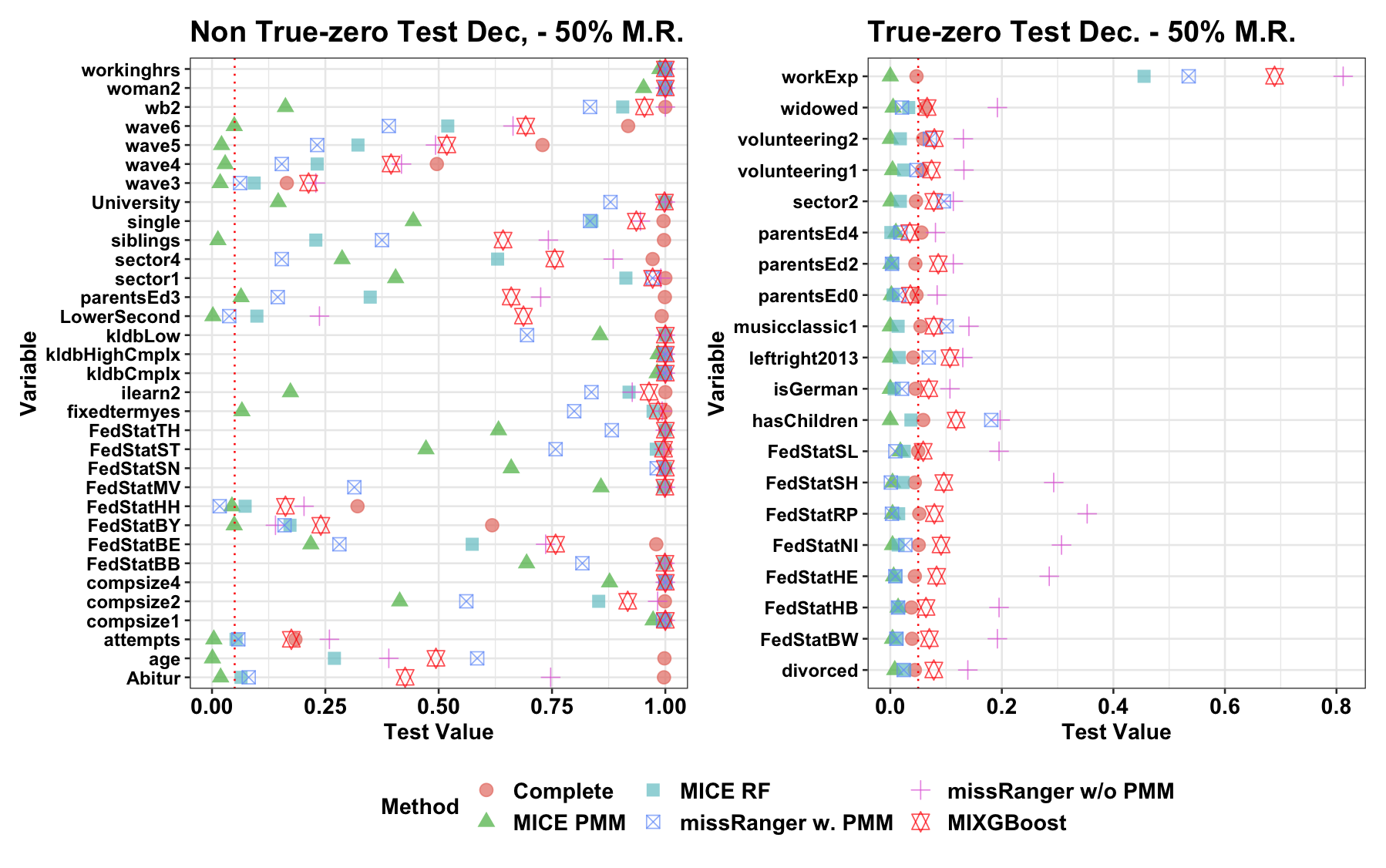}
 \end{subfigure}
\begin{minipage}{\textwidth}
{\scriptsize{ \textit{Note:} The Figure shows the rejection rates of the true coefficient, with higher values indicating a higher rejection rate. On the left for coefficients not equal to zero (as fixed in Equation \ref{eq1}), and on the right for coefficients equal to zero. "Complete" refers to estimation before any data are amputated, and "Deleted" refers to regression on listwise deleted datasets. The others refer to the imputation methods explained in the text.}}
\end{minipage}
\end{figure}

The $H_0$ rejection rates for $50\%$ missingness, as shown in Subfigure \ref{fig:rejection50} and aggregated in Table \ref{decaveragezero}, suggest (unsurprisingly) more drastic errors. For statistical power (non-zero coefficients, left plot), MICE PMM is again too conservative and almost never reaches the  desired statistical power. Comparing the mean and median for MICE PMM shows that only for this methods the median rejection rate is worse than the mean rejection rate, suggesting that some coefficient estimates improve the summary, while for the other methods some coefficient estimates make the whole summary look worse. The mean rejection rates of $0.6781$, $0.7854$, $0.5850$ and $0.7805$ for MICE RF, missRanger without PMM,  missRanger with PMM, and MIXGBoost show slightly better performance, but still fall short of the desired power. For all three methods, the median is again higher than the mean, indicating that there are some unfavorable outliers. I.e., MICE RF still rejects more than half of the coefficients at least $90\%$ of the time, but has a high variation (standard deviation of $SD=0.3728$), and missRanger without PMM and MIXGBoost reject most of the coefficients at least $95\%$ of the time (median $=0. 9840$ and median $=0.9590$, respectively) and remain relatively consistent ($SD=0.2971$ and $SD=0.2858$, respectively). missRanger with PMM fails to reach the desired threshold more often (median $=0.7265$) and is underpowered compared to MICE RF, missRanger without PMM, and MIXGBoost (Table \ref{decaveragezero} Panel B).  
In general, Subfigure \ref{fig:devTDMean} and \ref{fig:devTDMedian} on the right show that the average power declines with the increase in missingness, while the median power is in general higher and more stable. 

To sum up,  missRanger with PMM seems to be the best compromise for 10\% and 30\% missingness
when taking both, Type I error control and power, into account. If the Type I error is most important, MICE RF is the next best candidate, while it is MIXGBoost if the power is most important. For 50\% missingness, MICE RF outperforms missRanger with PMM.

\subsection{Imputation Metric and Computation Times}\noindent
Figure \ref{fig:ipm} in the Appendix shows the results of the IPM Metric and the Computational (running) times for all the methods and missingness rates evaluated in this study. Of these, MICE RF has the best IPM result followed by MIXGBoost, which means that these imputation methods are the closest to the true dataset (measured in IPM), but they are also the ones that take the most time to compute. MICE PMM, on the other hand, is the fastest method while showing a good IPM result. The IPM results for both missRanger methods show a clear difference between the 10\% and 50\% missingness, which could be a reason why missRanger with PMM performs worse with an increase in the missingness. 

\begin{table}[hbt!]
\caption{Aggregated Rejection Rates of all Methods and Missingness Rates} \label{decaverage}
\centering
\scriptsize
\begin{tabularx}{\textwidth}{Xccccccccc}
\toprule
& \multicolumn{3}{c}{\textbf{10\% Missingness}} & \multicolumn{3}{c}{\textbf{30\% Missingness}} & \multicolumn{3}{c}{\textbf{50\% Missingness}} \\ 
\cmidrule(lr){2-4} \cmidrule(lr){5-7} \cmidrule(lr){8-10}

  & Mean & Median & SD & Mean & Median & SD & Mean & Median & SD\\
  & (1) & (2) & (3) & (4) & (5) & (6) & (7) & (8) & (9)\\
\midrule

\multicolumn{4}{l}{\textbf{Panel A: Type I Error}} \\ \cmidrule{1-2}
Complete data & 0.0495 & 0.0490 & 0.0068 & 0.0506 & 0.0515 & 0.0060 & 0.0490 & 0.0470 & 0.0071\\
List-wise deletion & 0.0683 & 0.0520 & 0.0395 & & & & & & \\
MICE PMM & 0.0080 & 0.0090 & 0.0050 & 0.0180 & 0.0195 & 0.0095 & 0.0043 & 0.0040 & 0.0050\\
MICE RF & 0.0271 & 0.0275 & 0.0050 & 0.0497 & 0.0440 & 0.0340 & 0.0386 & 0.0155 & 0.0985\\
missRanger w/o PMM & 0.0640 & 0.0640 & 0.0122 & 0.1032 & 0.0865 & 0.0432 & 0.2096 & 0.1665 & 0.1617\\
missRanger with PMM & 0.0385 & 0.0340 & 0.0165 & 0.0464 & 0.0355 & 0.0546 & 0.0639 & 0.0210 & 0.1197\\
MIXGBoost & 0.0520 & 0.0540 & 0.0093 & 0.0617 & 0.0580 & 0.0260 & 0.1067 & 0.0780 & 0.1385\\
\midrule
\multicolumn{4}{l}{\textbf{Panel B: Statistical Power}} \\ \cmidrule{1-2}
Complete data & 0.8937 & 1.0000 & 0.2403 & 0.8898 & 0.9995 & 0.2419 & 0.8930 & 1.0000 & 0.2410\\
List-wise deletion & 0.5598 & 0.5290 & 0.3286 & & & & & &  \\
MICE PMM & 0.5722 & 0.6255 & 0.3999 & 0.5545 & 0.6520 & 0.3998 & 0.3987 & 0.2525 & 0.3850\\
MICE RF & 0.7771 & 0.8985 & 0.3059 & 0.6981 & 0.8470 & 0.3295 & 0.6781 & 0.9095 & 0.3754\\
missRanger w/o PMM & 0.8759 & 0.9990 & 0.2570 & 0.8300 & 0.9740 & 0.2756 & 0.7854 & 0.9865 & 0.2971\\
missRanger with PMM & 0.8554 & 0.9925 & 0.2754 & 0.7159 & 0.8785 & 0.3312 & 0.5850 & 0.7265 & 0.3716\\
MIXGBoost & 0.8830 & 0.9980 & 0.2494 & 0.8425 & 0.9740 & 0.2628 & 0.7805 & 0.9590 & 0.2858\\
\bottomrule
\end{tabularx}  
\begin{minipage}{\textwidth}
{\scriptsize{  \textit{Note:} We counted whether the test was rejected or not for each replication and then took the mean and median rejection rates of separated by variables with a true zero coefficient, showing the type I error rate in Panel A and C. We did the same for non true-zero coefficients in Panel B and D showing the statistical power. This simplifies Figure \ref{fig:rejection1}. The standard deviation shows how much the test decision varies.}}
\label{decaveragezero}
\end{minipage}
\end{table}

\section{Discussion}\noindent
The goal of the simulation study was to evaluate how tree-based imputation methods compare to each other and to MICE PMM. MICE PMM is, up to now, the gold standard in social science research for imputing missing data \parencite{CostantiniEA2023}. As tree-based methods, we used three Random forest based imputation methods, MICE with Random Forest (RF) as well as missRanger without and with PMM, and MIXGBoost, an imputation method based upon extremely gradient boosting. Our evaluation of the different imputation methods revealed certain performance patterns as summarized below.

\textit{Coefficient Estimation after Imputation.} The biases were higher for non true-zero coefficients than for true-zero coefficients and generally low for all methods. At 10\% missing data, MICE PMM had the highest bias, while the tree-based methods had a comparable low bias, with MIXGBoost having the lowest bias overall. Interestingly, the bias for MICE PMM decreased with an increase of the percentage of missing data up to 50\%, having a similar bias to missRanger without PMM. missRanger with PMM had the highest bias with increased missingness, while MIXGBoost had the lowest. 

Although the bias was generally very low, we seldom observed a situation where the estimated coefficient changed direction when using multiple imputation (estimating a positive coefficient when the true coefficient is negative) in some of the replications. This seldom fact was particularly the case for missRanger when missingness was high. To sum up, all methods performed reasonably well in terms of coefficient estimation, which was also seen in the empirical example of \textcite{ZeyerETunderreview}, where most of the estimates of the imputation methods included in that study were close to each other.

\textit{Hypothesis Testing: Type I Error and Power after Imputation.} In terms of Type I and statistical power, MICE PMM proved to be the most conservative, with very low Type I error rates, well below the .05 threshold, and very low statistical power. Thus, although it is very unlikely to falsely reject a true zero, it is also not likely to rightfully reject non-true zero coefficients with MICE PMM. Therefore, empirical studies using MICE PMM to impute missing values may fail to find true significant effects due to its conservative nature. Conversely, missRanger without PMM, together with MIXGBoost, had the highest power, but exhibit an unsatisfactory Type I Error control with too liberal true levels that were much larger than the prescribed 5\% (e.g., more than 20\% on average for 50\% missingness). The Type I error of MIXGBoost was better, but still exhibited rather large rejection rates around 10\%  for 50\% missingess.

In our study, MICE RF followed by missRanger with PMM and MIXGBoost show the overall best results. While MICE RF is always below the 5\% threshold, missRanger with PMM exceeds this threshold only sharply for 50\% missingness. In addition, MICE RF and missRanger with PMM have much higher power than MICE PMM. The power is higher for missRanger with PMM for 10\% missingness, while it was higher for MICE RF for 50\% missingness. In case of 30\% missingness, the power is similar. This can also be seen in the empirical study by \textcite{ZeyerETunderreview}, who found that coefficients were rejected as unequal to zero more often when imputing data with missRanger with PMM than with MICE PMM. For MIXGBoost, while the power is on average better than for MICE RF and missRanger with PMM, the Type I error is much worse (above 0.10 on average) for 50\% missingness.

\textit{Imputation Accuracy measured by IPM.}
MICE RF and MIXGBoost provide the best imputed prediction mean (IPM) results, followed by missRanger with PMM, but only for 10\% missingness. At 50\% missingness, MICE PMM has the better IPM metric. This may explain why missRanger with PMM performs worse with increasing missingness. For MICE RF, however, the IPM metric are much closer, but performance also slightly decreases as missingness increases.

While MICE RF and MIXGBoost appears to be the best imputation method with respect to the IPM metric in this setting, it comes at the cost of high computational cost. With run times ranging from 18 to 26 minutes, it took much longer than MICE PMM (4 to 7 minutes) and missRanger without (5 to 10 minutes) and with (6 to 15 minutes) PMM, which are much closer. The running time of MICE RF could become a problem with a larger sample sizes or in high-dimensional settings (where the number of variables is close to or greater than the number of observations). Considering the overall trade-offs between accuracy and rejection rates, MICE RF seems to provide the best balance, with missRanger close behind, especially for lower missingness rates. 

Ultimately, the choice between these (or any) imputation methods may depend on the specific characteristics of the dataset, the research question, and the resources available. The results of this study show that bias is generally low. Still, MIXGboost had the lowest overall bias, but MICE PMM and MICE RF had lower bias for true zero coefficients. When test decisions are of high importance to the research question, MICE RF and missRanger with PMM have (mostly) good Type I error rates, while also having much higher power than MICE PMM. If computationally feasible, we recommend MICE RF, and if not, missRanger with PMM.

\subsection{Limitations \& Outlook}\noindent
Our study design has some limitations that we must acknowledge. First and foremost, although tree-based methods shine on high-dimensional data, our simulated data were constructed in close relation to a social science research question that typically has a rather large number of regressions but not where the number of variables is close to or greater than the number of observations (high-dimensional data). This limits the generalizability of our results to more complex datasets. In addition, the simulation design predominantly involves linear relationships, which may overlook the nuances and subtleties of nonlinear dynamics in real-world data. This may affect the robustness of our conclusions, especially with respect to the performance of imputation methods in datasets characterized by complicated relationships and interactions. 

However, even in our situation, the tree-based methods, especially MICE RF and missRanger with PMM, are clearly superior to MICE PMM, but the latter only if the missingness rate is not too high. \textcite{AkandeLR2017} and \textcite{Murray2018} have already shown that MICE with Classification and Regression Trees (CART) can outperform MICE PMM for multiple imputation of categorical data. However, MICE CART is much slower than MICE RF, rather susceptible to overfitting and first tests have shown a superiority of MICE RF. To reduce the output, MICE CART is therefore not included in our simulation study.

Future research on this topic could explore intermediate levels of missingness to get a better understanding when the performance of missRanger with PMM decreases. In addition, more contextual information or more context for the imputation itself could be collected to find guidance on which imputation method performs best in which settings. Furthermore, the analysis steps could be extended to panel regression methods or to multilevel data which is particularly difficult to impute \parencite{EndersT2007, LudtkeRG2017, grundluedtkesimulationsandreccomendations}. Finally, another interesting avenue might be sensitivity analyses on subsets 
, where the sample is divided based on different variables, such as states, to highlight whether in the empirical analysis one should first subset the data for the final regression analysis or apply imputation first.

\printbibliography

\appendix

\section{Distribution comparisons}\label{sec:A1}

Along this subsection and in order to display together the original distribution of the variables and the simulations, plots with the distribution of each feature are shown for both cases. In each graph, the original NEPS data is displayed with bars filled with blue color and the simulated data for seed 7 is displayed as unfilled red bars.

\begin{figure}[ht!]
\caption{Comparing NEPS and simulated distributions}
    \begin{subfigure}{\textwidth}
        \centering
     \subcaption{Distribution comparisons: Constant independent features} \label{dplotsA}
     \includegraphics[width=0.475\textwidth]{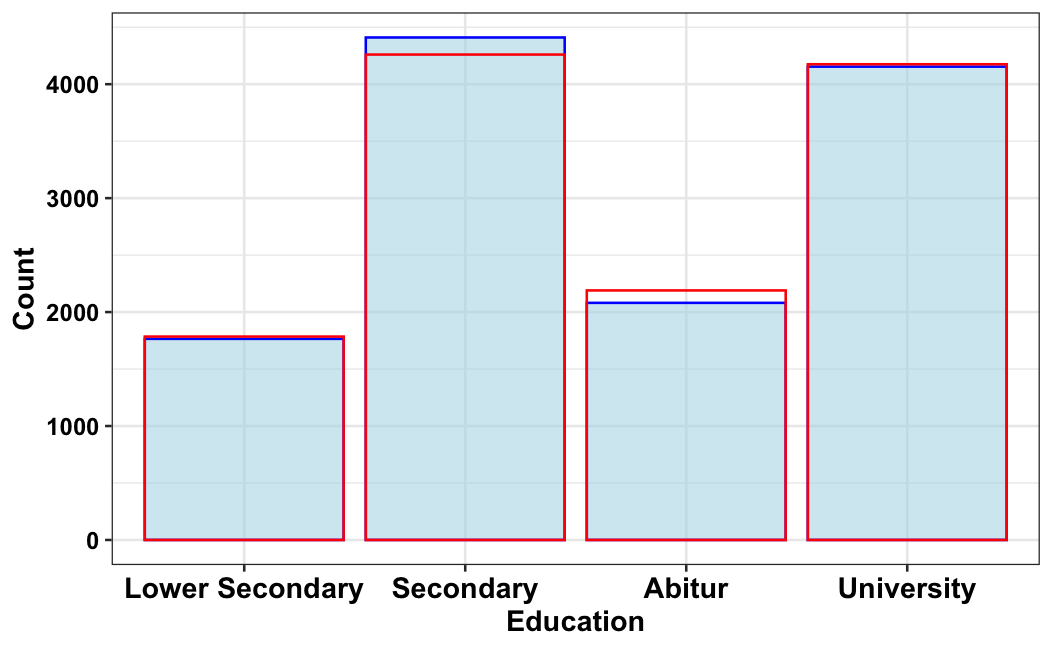} \hfill
        \includegraphics[width=0.475\textwidth]{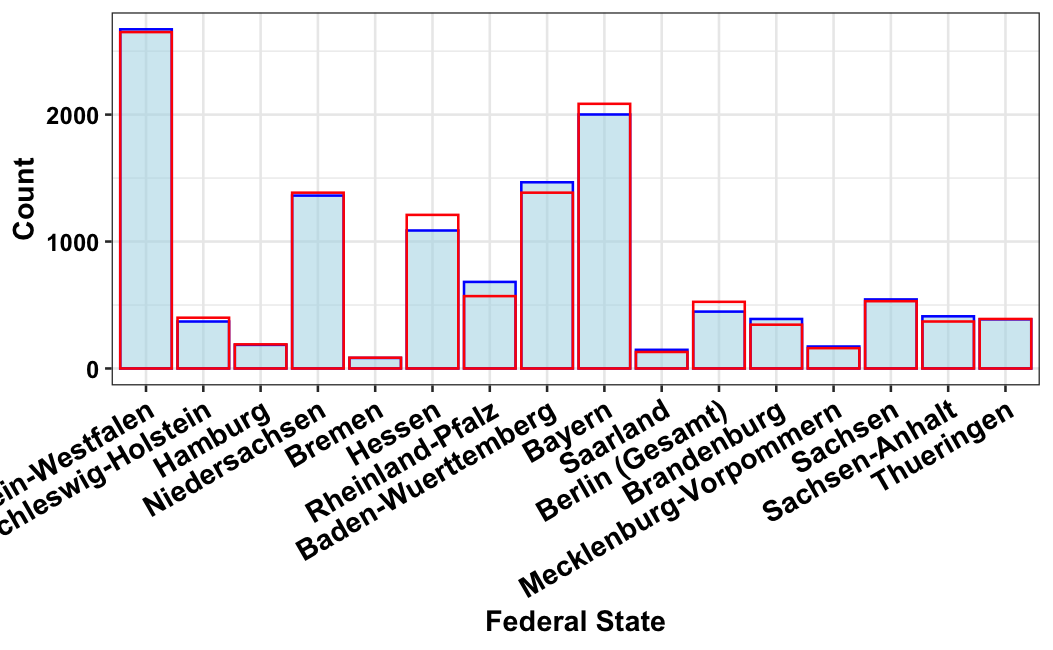} \\
     \includegraphics[width=0.475\textwidth]{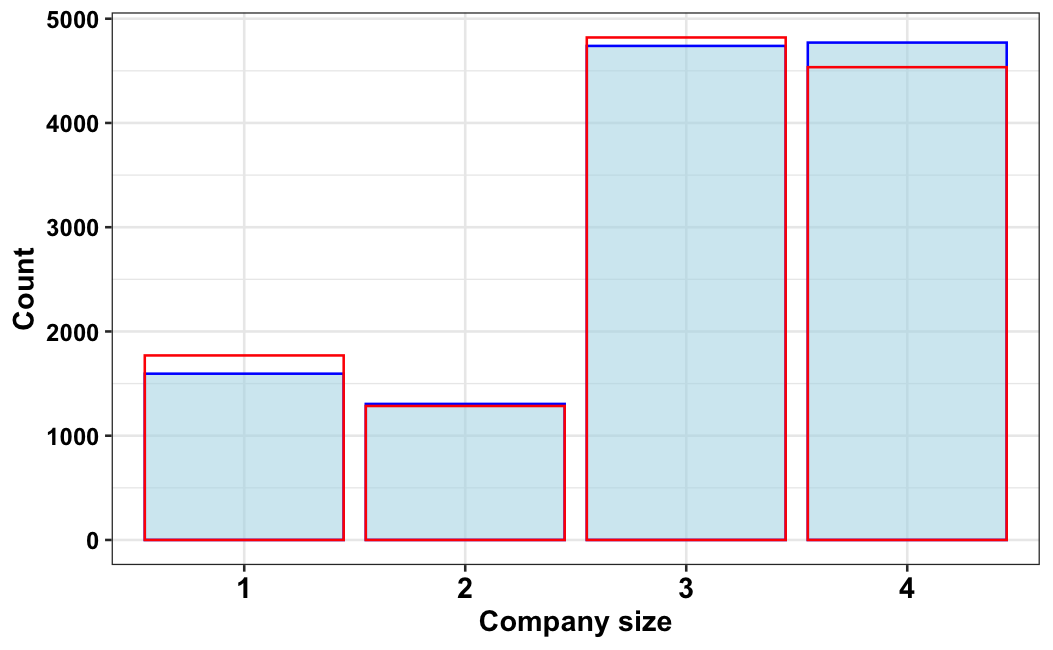} \hfill
         \includegraphics[width=0.475\textwidth]{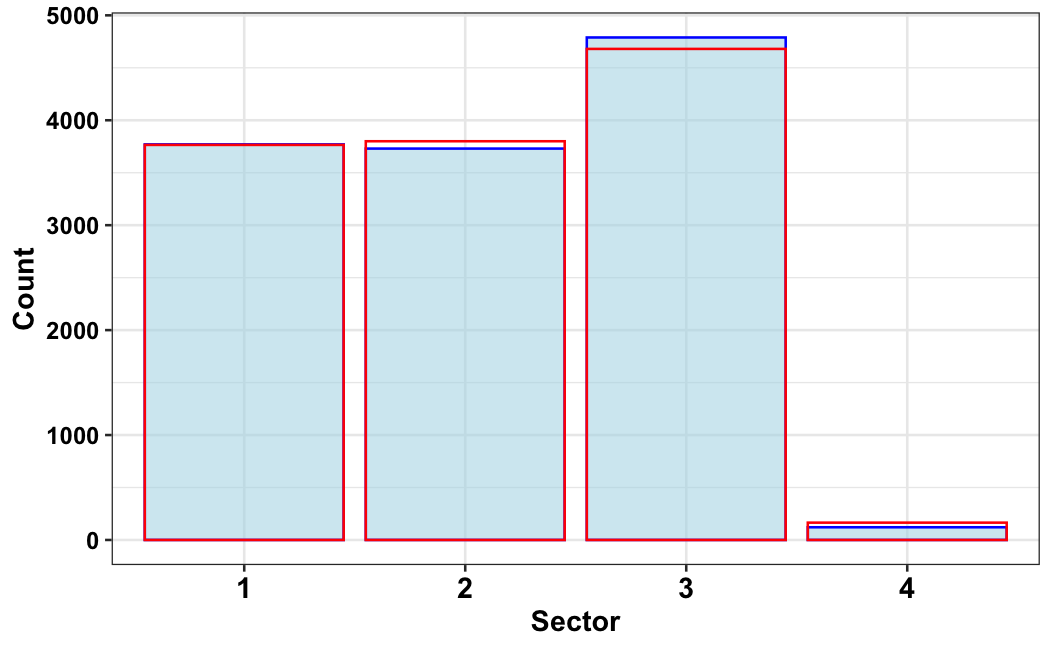} \\
     \includegraphics[width=0.475\textwidth]{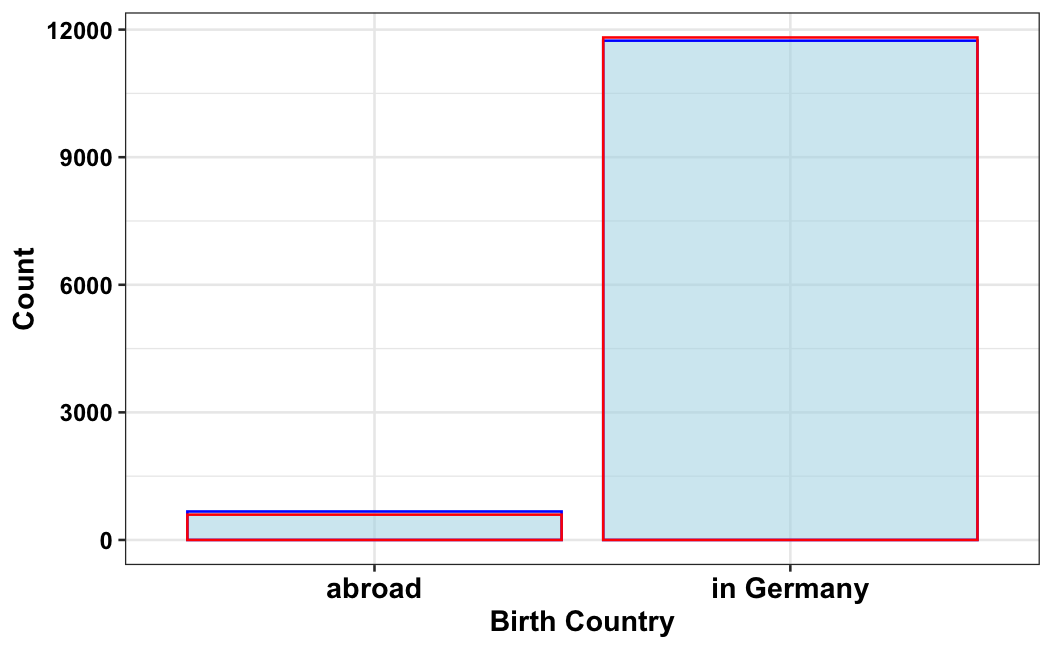} \hfill
        \includegraphics[width=0.475\textwidth]{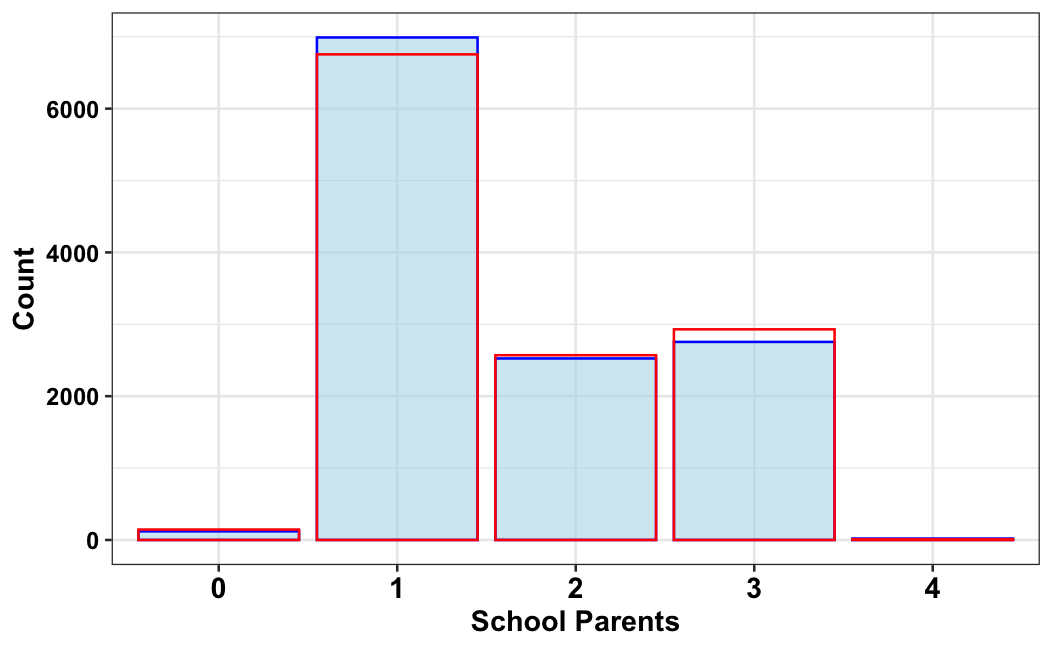} \\
        \end{subfigure}
\end{figure} 

\begin{figure}[ht!]
\ContinuedFloat
    \begin{subfigure}{\textwidth}
        \centering
     \includegraphics[width=0.475\textwidth]{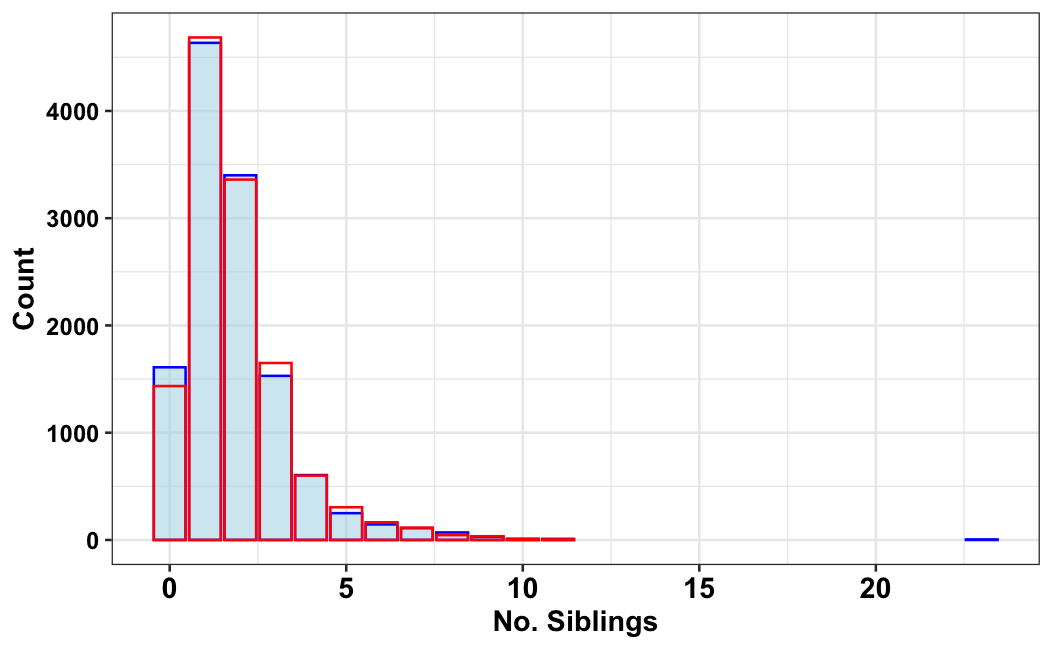} \hfill
        \includegraphics[width=0.475\textwidth]{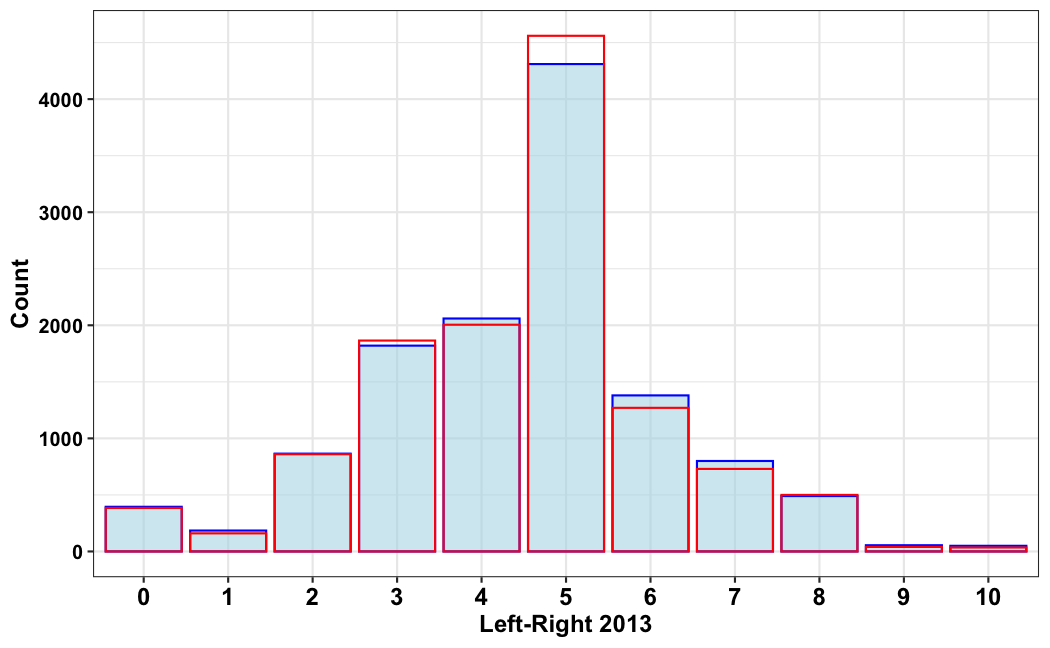} \\
     \includegraphics[width=0.475\textwidth]{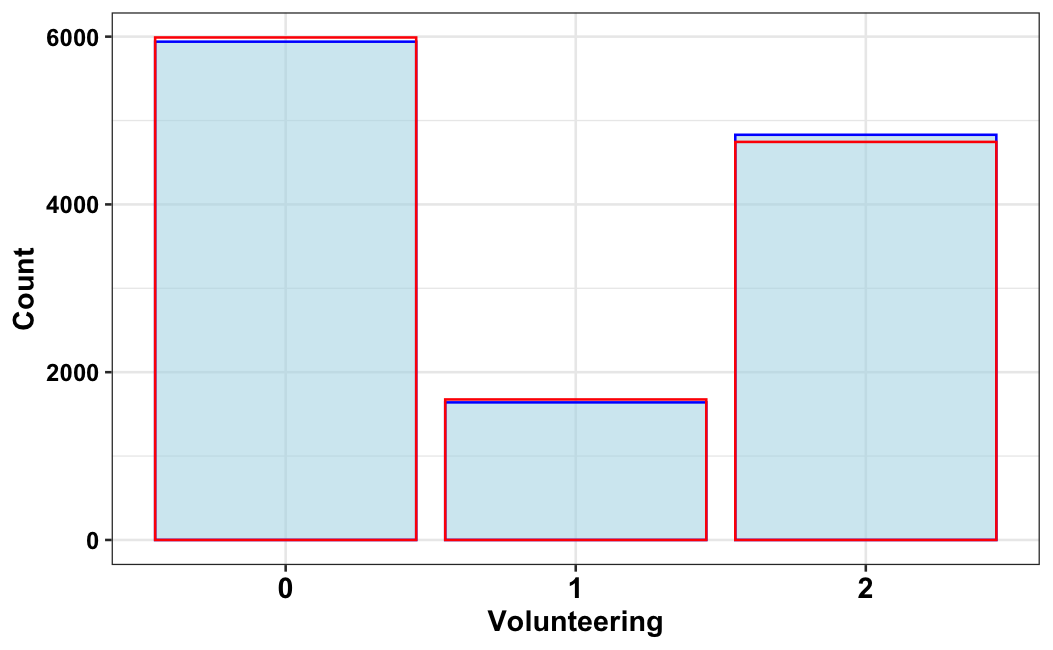} \hfill
        \includegraphics[width=0.475\textwidth]{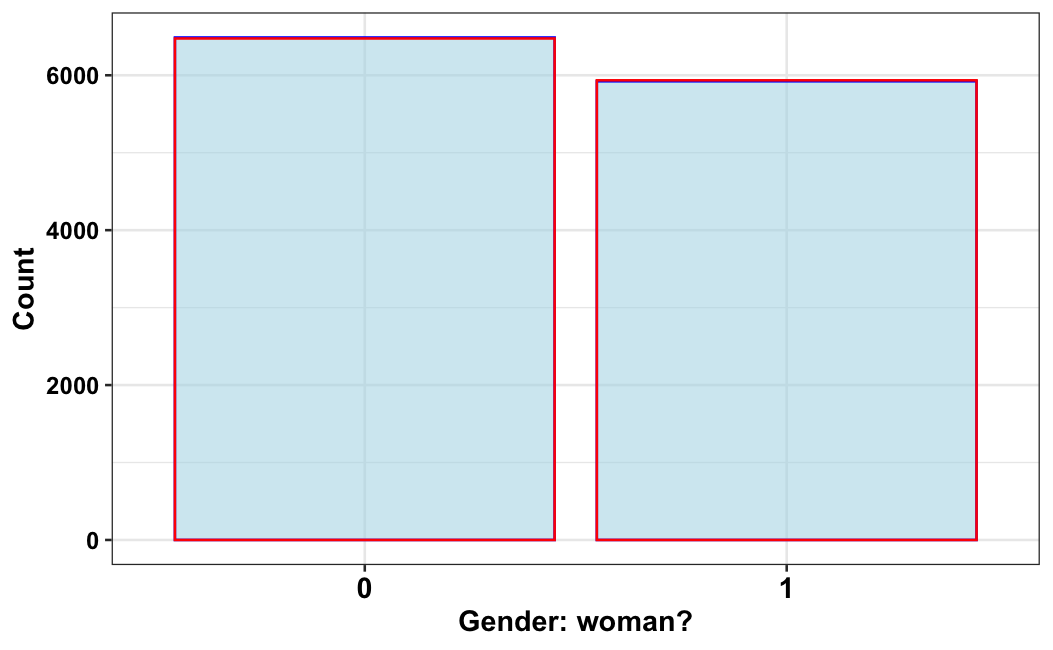} \\
     \includegraphics[width=0.475\textwidth]{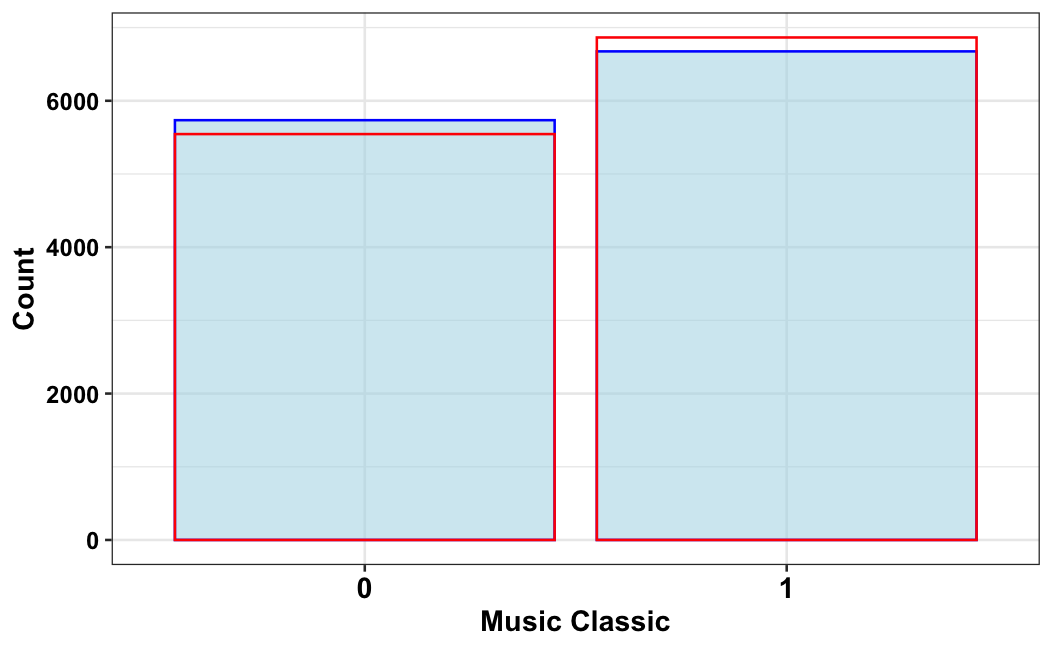} \hfill
     \end{subfigure}
\end{figure} 

\begin{figure}[ht!]
\ContinuedFloat
    \begin{subfigure}{\textwidth}
        \centering
     \subcaption{Distribution comparisons: Time-varying independent features} \label{dplotsB}
     \includegraphics[width=0.475\textwidth]{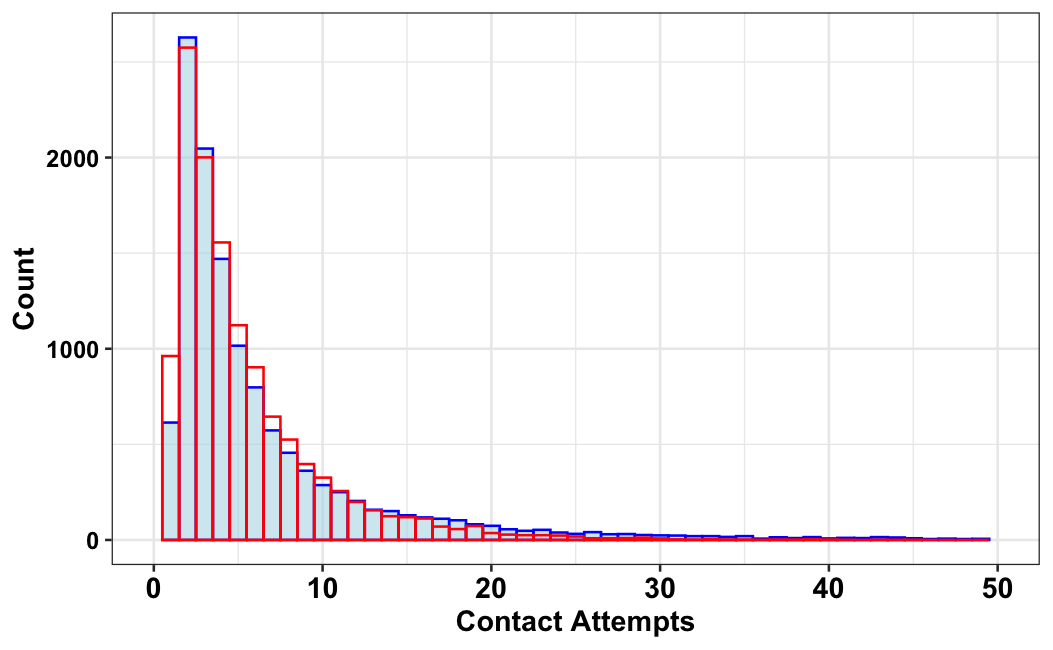} \hfill
        \includegraphics[width=0.5\textwidth]{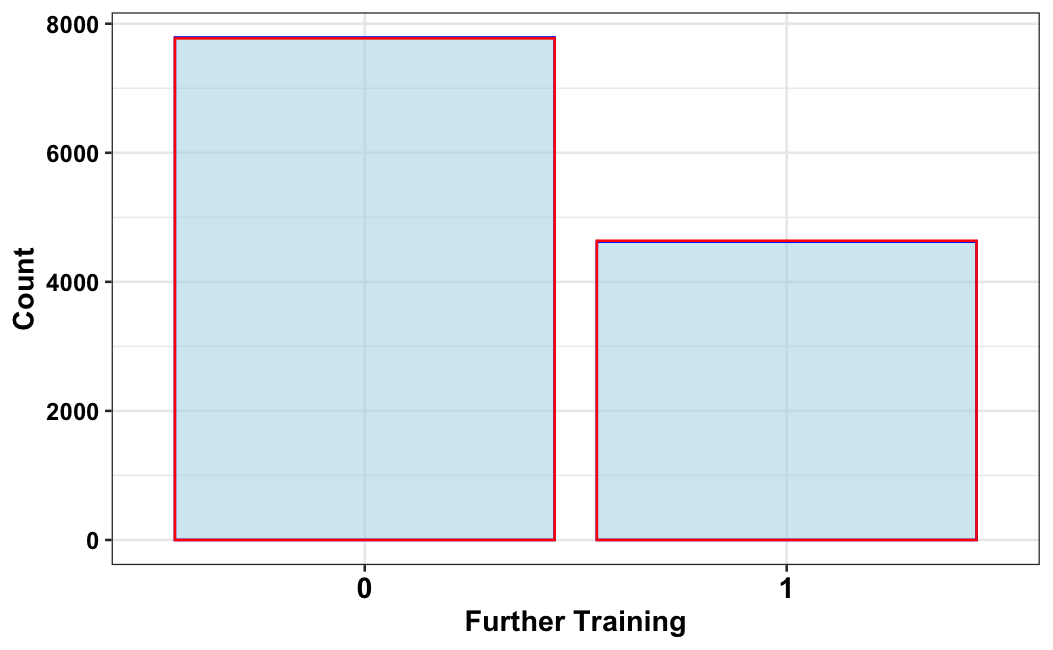} \\
     \includegraphics[width=0.475\textwidth]{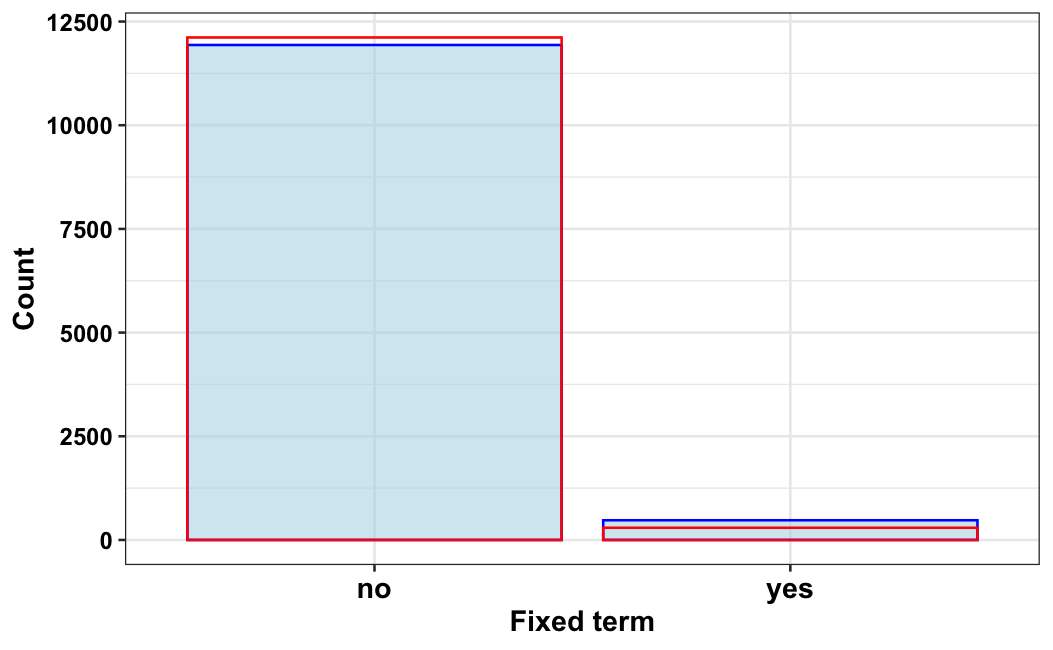} \hfill
        \includegraphics[width=0.475\textwidth]{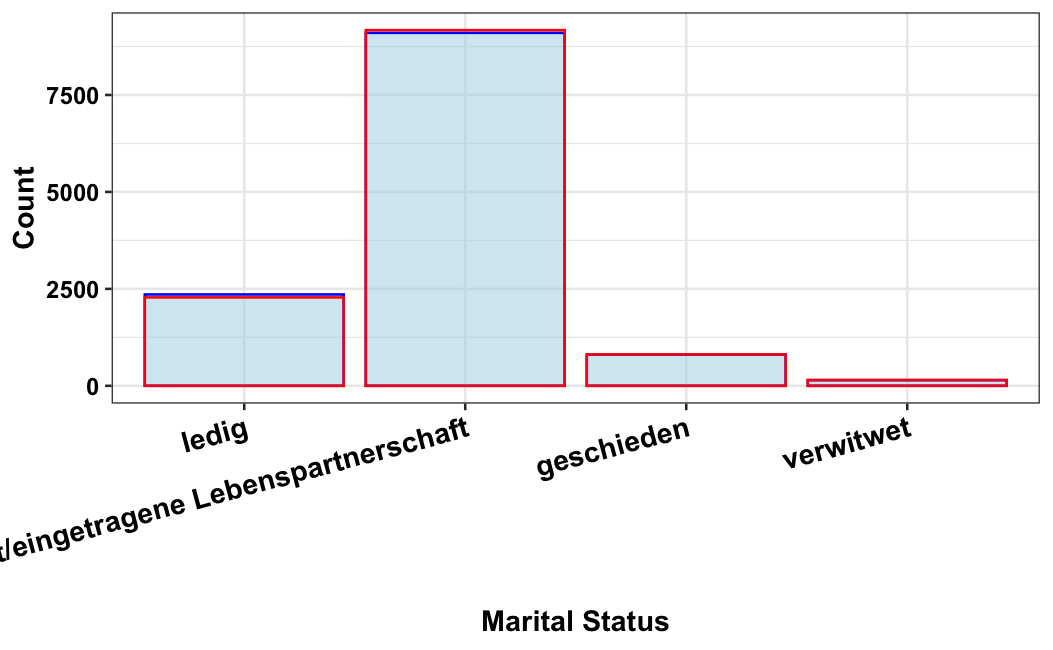} \\
     \includegraphics[width=0.475\textwidth]{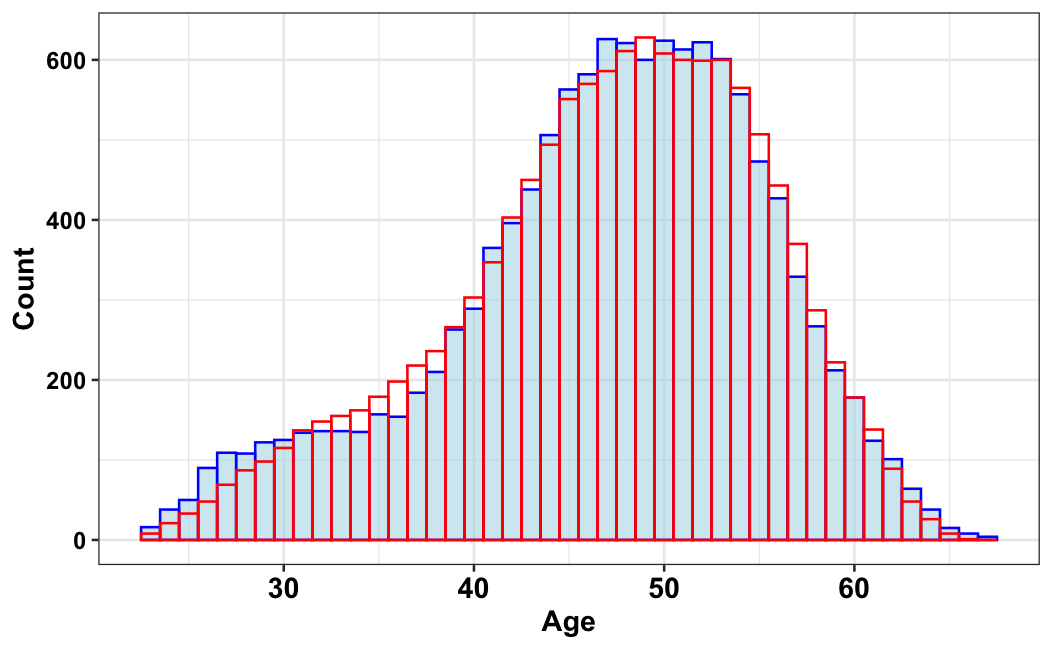} \hfill
        \includegraphics[width=0.475\textwidth]{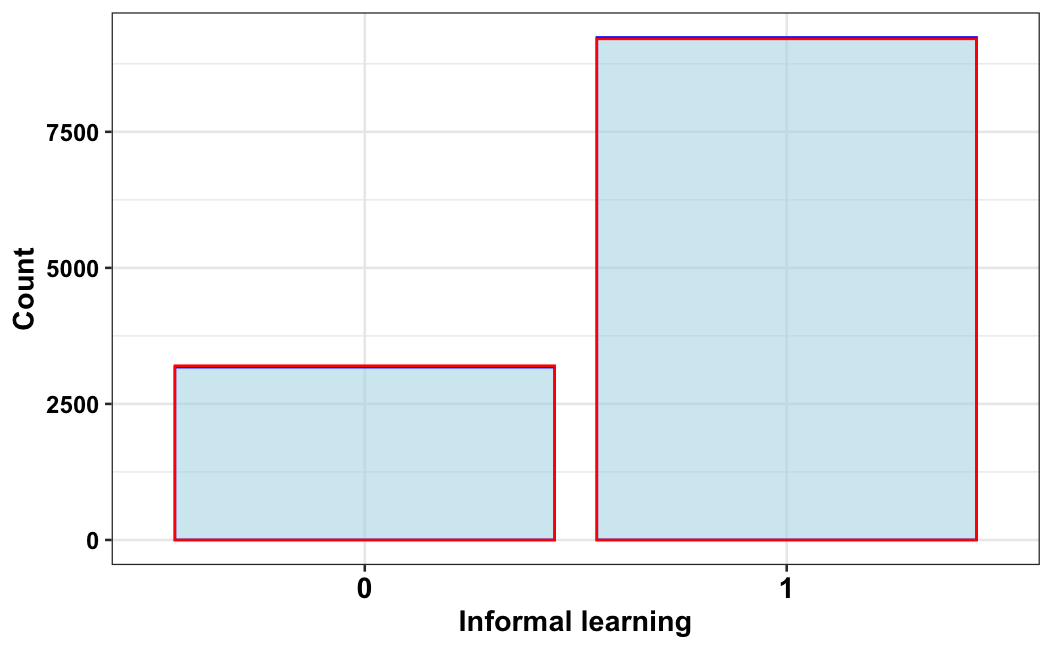}
        \end{subfigure}
\end{figure} 

\begin{figure}[ht!]
\ContinuedFloat
    \begin{subfigure}{\textwidth}
        \centering
     \subcaption{Distribution comparisons: Constant endogenous features} \label{dplotsC}
     \includegraphics[width=0.475\textwidth]{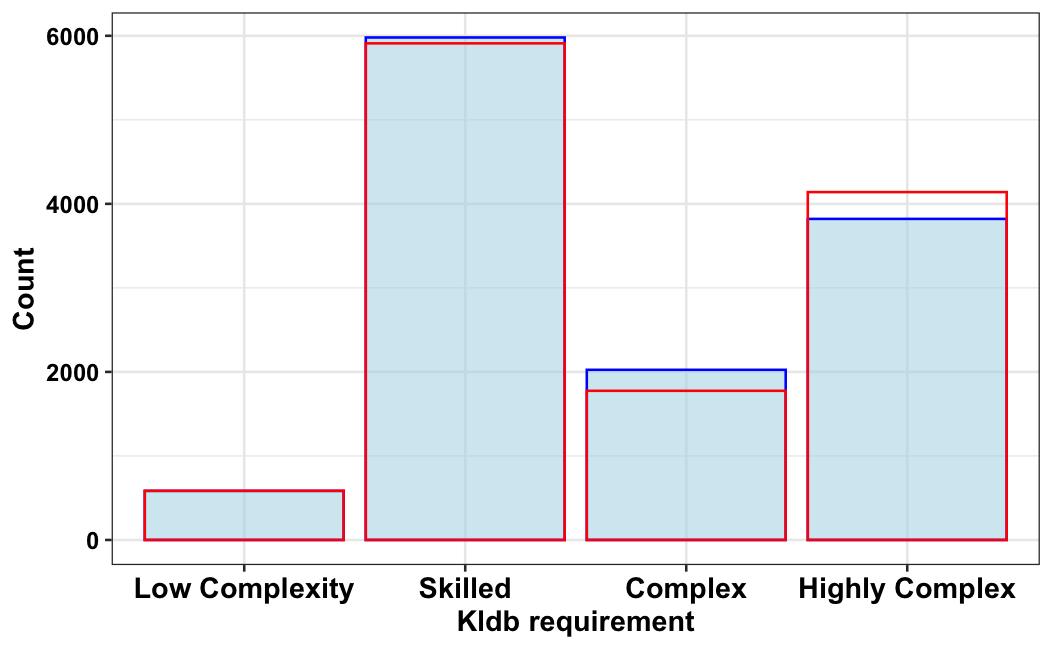} \hfill
        \includegraphics[width=0.475\textwidth]{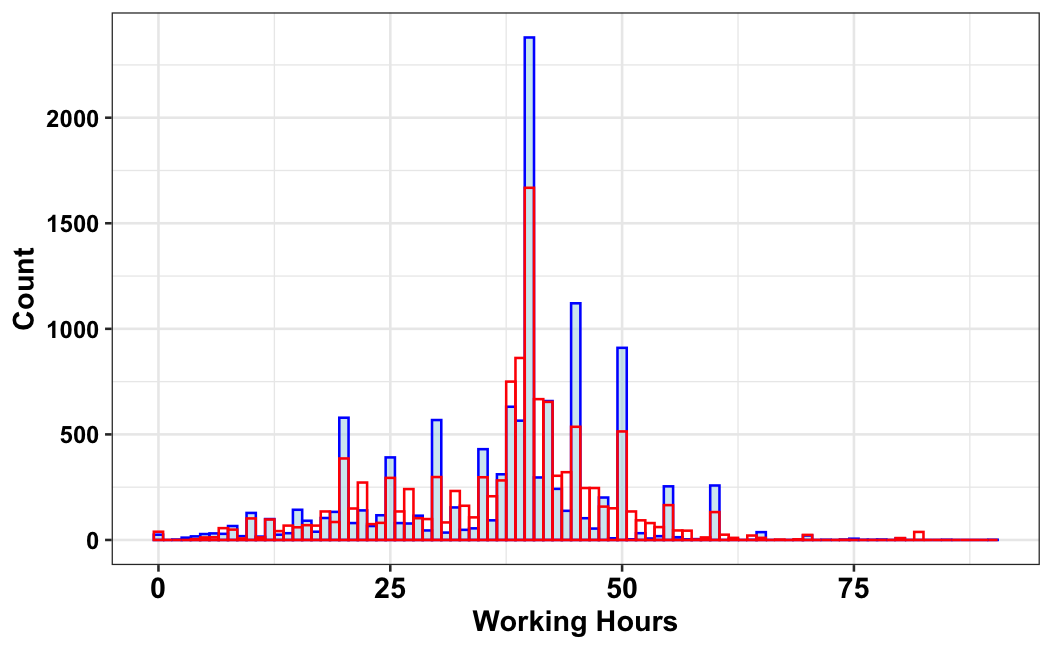}
        \end{subfigure}
\end{figure} 

\begin{figure}[ht!]
\ContinuedFloat
    \begin{subfigure}{\textwidth}
        \centering
     \subcaption{Distribution comparisons: Time-varying endogenous features} \label{dplotsD}
     \includegraphics[width=0.475\textwidth]{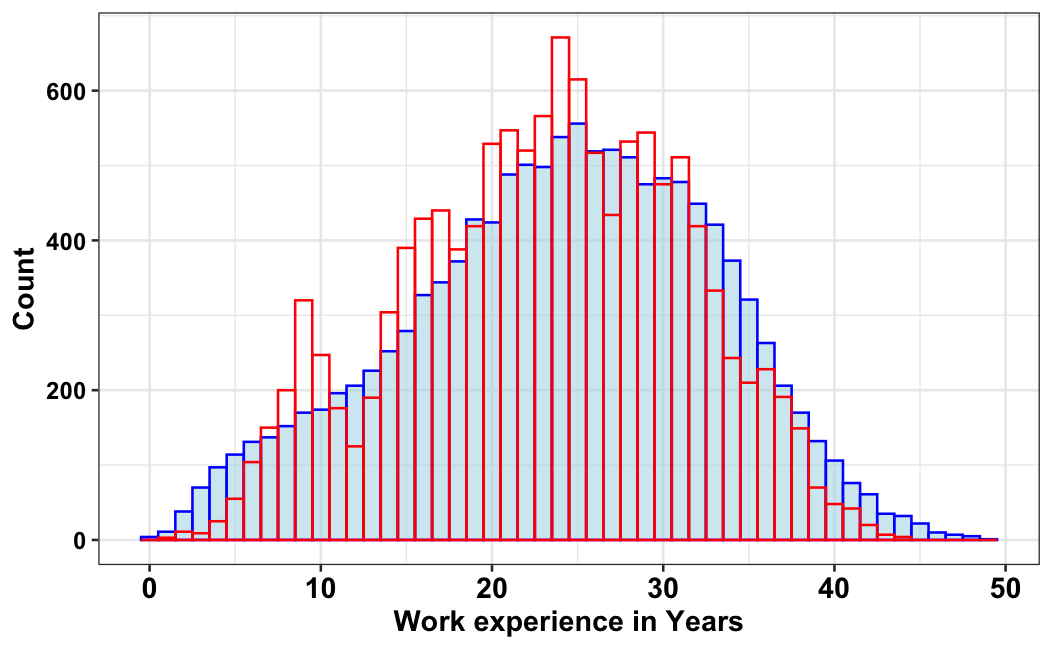} \hfill
        \includegraphics[width=0.475\textwidth]{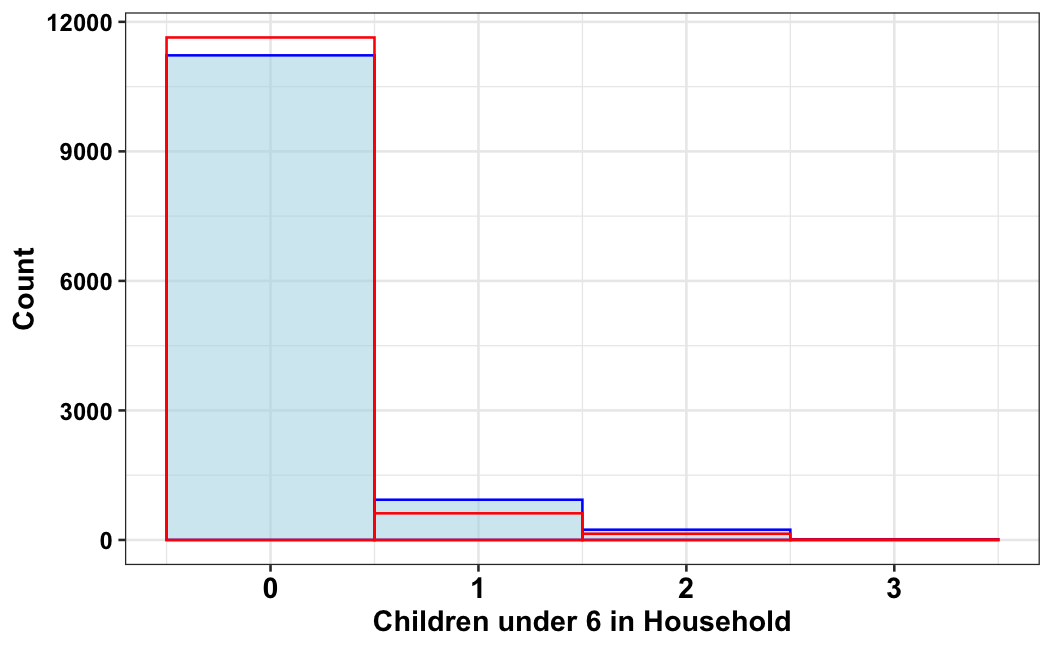} \\
\end{subfigure}
\end{figure} 

\begin{figure}[ht!]
\ContinuedFloat
    \centering    \begin{subfigure}{\textwidth}
        \centering     \subcaption{Distribution comparisons: Outcome variable} \label{dplotsE}
     \includegraphics[width=0.475\textwidth]{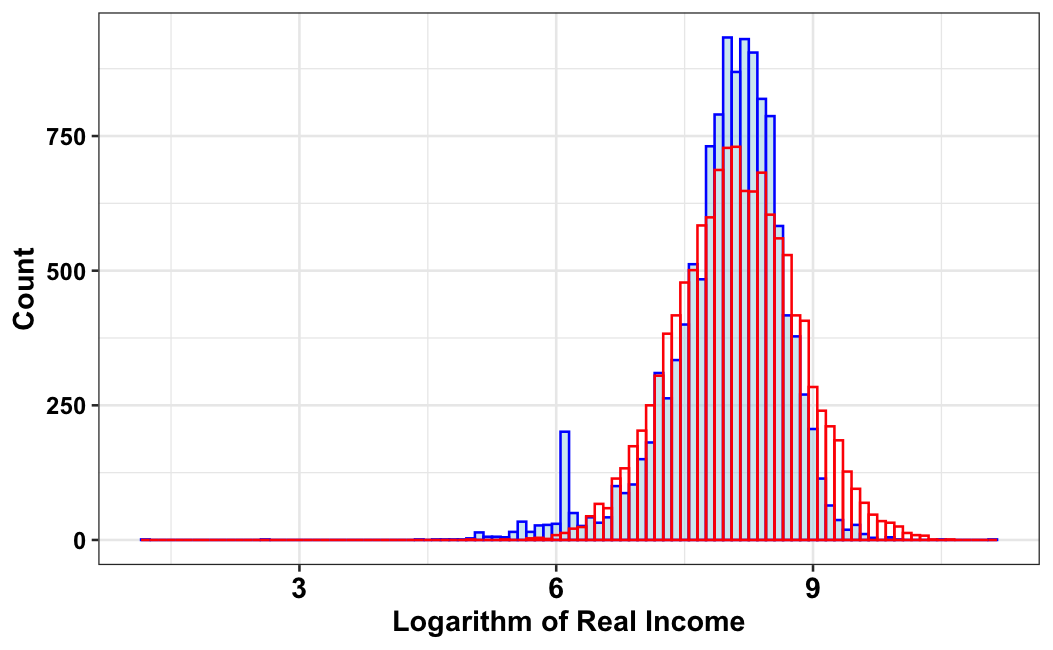} \\
\end{subfigure}
\end{figure}

\clearpage \newpage

\section{Description of the Missing Algorithm}\label{sec:A2}

Let us consider a dataset with n observations measured on p variables given by 
$Y= (y_{i,j})_{i,j}$ consisting of missing parts $Y_{mis}$ and observed parts $Y_{obs}$. Then the 
total missing rate $\nu$ is define by 
$$\nu = \frac{|Y_{mis}|}{|Y_{mis}|+|Y_{obs}|},$$
where $|\cdot|$ is the number of entries under each group according to whether they are observed or missing. Let $\mathcal{M}$ also be a matrix full of zeros with the same dimensions as $Y$, which will be used to denote a missing entry with 1 and an observed one with 0. In our case, $\mathcal{M}$ is assumed to have no missing entries in the beginning, since we start with a complete dataset. For ease of notation, all columns of $Y$ that denote the $j$-th variable will be referred to simply as $Y_j$ from now on. First, one of the variables $Y_j \in \{Y_1, ..., Y_p\}$ is selected and completely missing entries are introduced at random with a total missingness rate $\nu$. At the same time, the matrix $\mathcal{M}$ is updated accordingly and the entries $\mathcal{M}_{ij}$ with missing values added to $Y_j$ are replaced by ones (1). After that, the following steps are repeated for the remaining variables, which are denoted as $Y_{-j} = \big( Y \setminus Y_j \big)$: 
\begin{itemize}
    \item Suppose $\mathbf{H} = [\eta^1, ..., \eta^m]$ is the vector containing the \textbf{not missing} unique values $\eta^\ell$ or distinct observations in $Y_j$ (with $\ell=1,\dots,m \leq n$), generate a probability vector $\mathbf{T}_j = [\tau^1_j,...,\tau^m_j]$ containing $M$ independent and identically distributed random numbers with $M$ denoting the number of entries in $\mathbf{H}$.  
    \item Sort $\eta^\ell$ in descending order and $\tau^\ell_j$ in ascending order and assign the largest $\eta^\ell$ to the smallest $\tau^\ell_j$, the second largest $\eta^\ell$ to the second smallest $\tau^\ell_j$ and so on until all $m$ assignments are made. These new assigned probabilities are now called $\mathbf{\hat{T}}_j = [\hat{\tau}^1_j,...,\hat{\tau}^\ell_j]$.
    \item Compute the vector of absolute frequencies of each $\eta^\ell$, denoted by $Z_j = [\zeta^1,...,\zeta^\ell]$.
    \item For each $\ell=1,...,m$ compute the probability of a missing value in $Y_{-j}$ given $\eta^\ell$ as follows: 
    $$\pi_\ell = \frac{\hat{\tau}^\ell_j \cdot \zeta^l}{\sum_{\ell=1}^m (\hat{\tau}^\ell_j \cdot \zeta^l)} \quad.$$
    \item Next, for each $\eta^\ell$, find the subset $\kappa^\ell$ of indexes $i=1,...,n$ for which the observation belongs to the group of $\kappa^\ell = \{i=1,...,n:y_{ij}=\eta^\ell\}$. That is, the subset of indices with value $\eta^\ell$ from the initially selected variable $Y_j$. 
    \item Randomly select $\hat{\nu} = \nu \cdot \pi_\ell$ entries from $\kappa^\ell$ and include a missing value in $Y_{-j}$ along with its corresponding update in $\mathcal{M}$, making $\mathcal{M}_{ij}=1$.
    \item Finally, repeat the above steps until you have finished with all $Y_{-j}$ variables.
\end{itemize} 

\clearpage \newpage

\begin{figure}[htbp!]

\caption{Development of the Bias over Missingness Rates}\label{fig:devBias}
    \begin{subfigure}{\textwidth}
        \centering
    \subcaption{Mean Bias} \label{fig:devBiasMean}
 \includegraphics[width=0.85\textwidth]{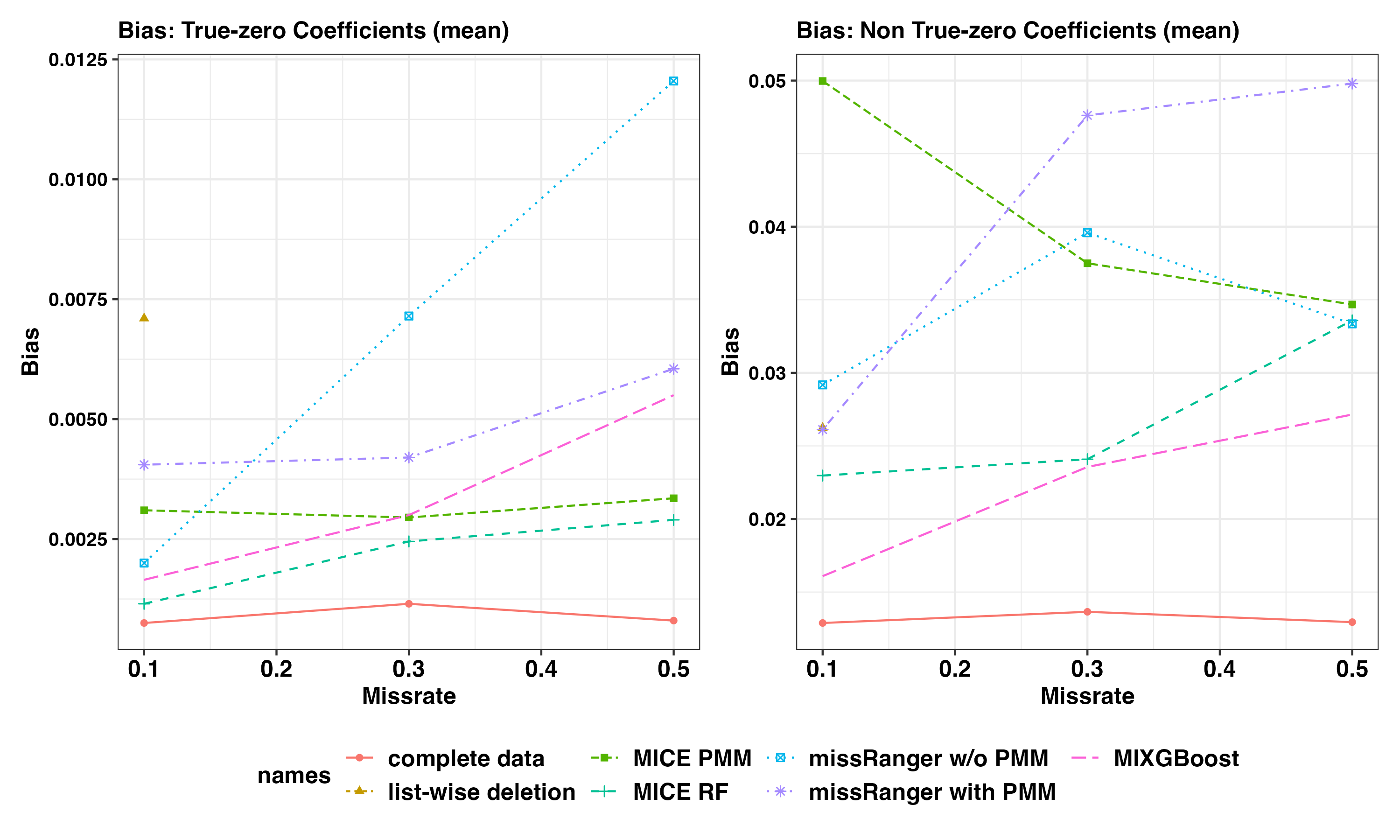}
 \end{subfigure}
     \begin{subfigure}{\textwidth}
        \centering
     \subcaption{Median Bias} \label{fig:devBiasMedian}
 \includegraphics[width=0.85\textwidth]{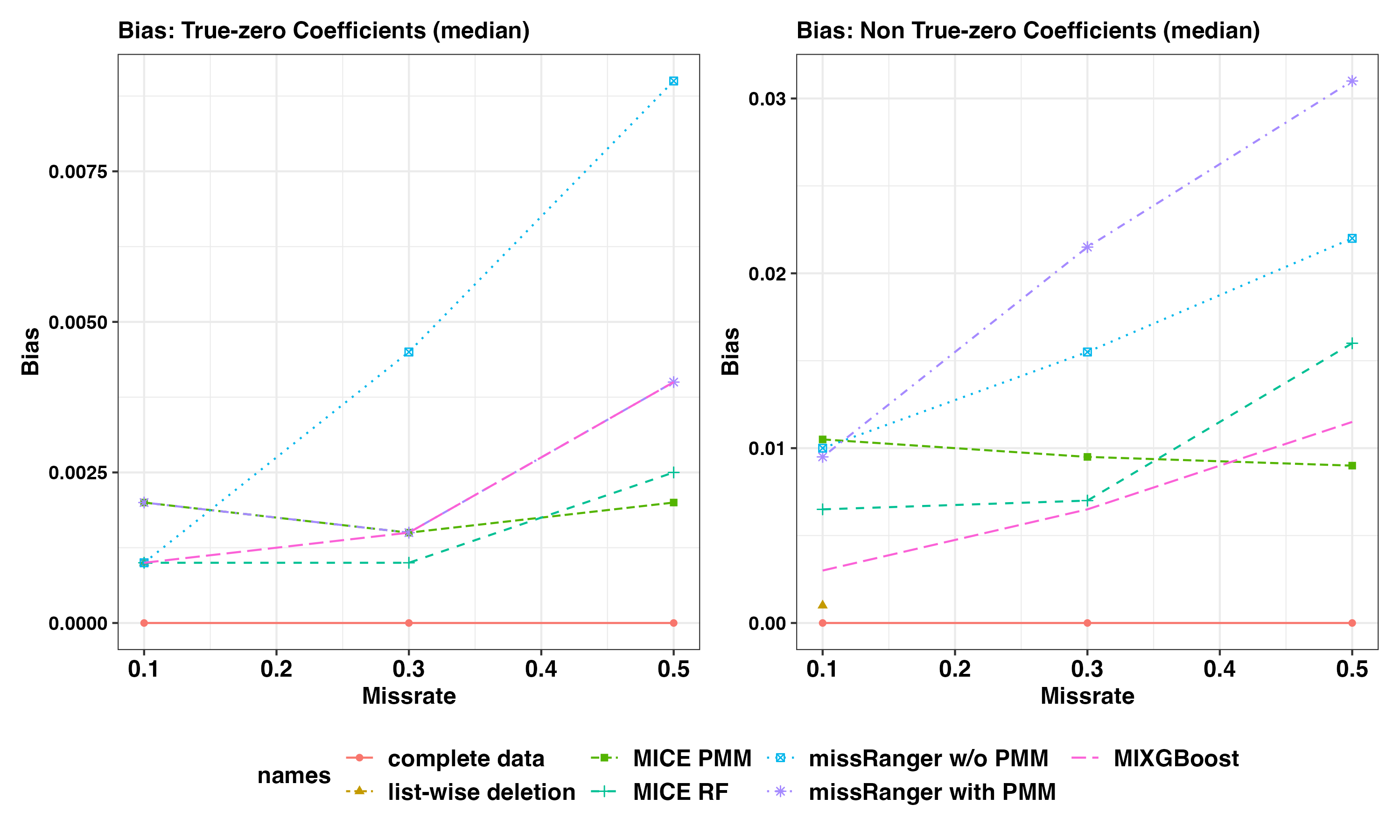}
  \end{subfigure}
  \caption*{\footnotesize{\textit{Note:} The Figure shows the rejection rates of the true coefficient, with higher values indicating a higher rejection rate. On the left for coefficients not equal to zero (as fixed in Equation \ref{eq1}), and on the right for coefficients equal to zero. "Complete" refers to estimation before any data are amputated, and "Deleted" refers to regression on listwise deleted datasets. The others refer to the imputation methods explained in the text.}}
\end{figure}

\begin{figure}[htbp!]

\caption{Development of the Rejection Rates over Missingness Rates}
\label{fig:devTD}
    \begin{subfigure}{\textwidth}
        \centering
    \subcaption{Mean Rejection Rates} \label{fig:devTDMean}
 \includegraphics[width=0.85\textwidth]{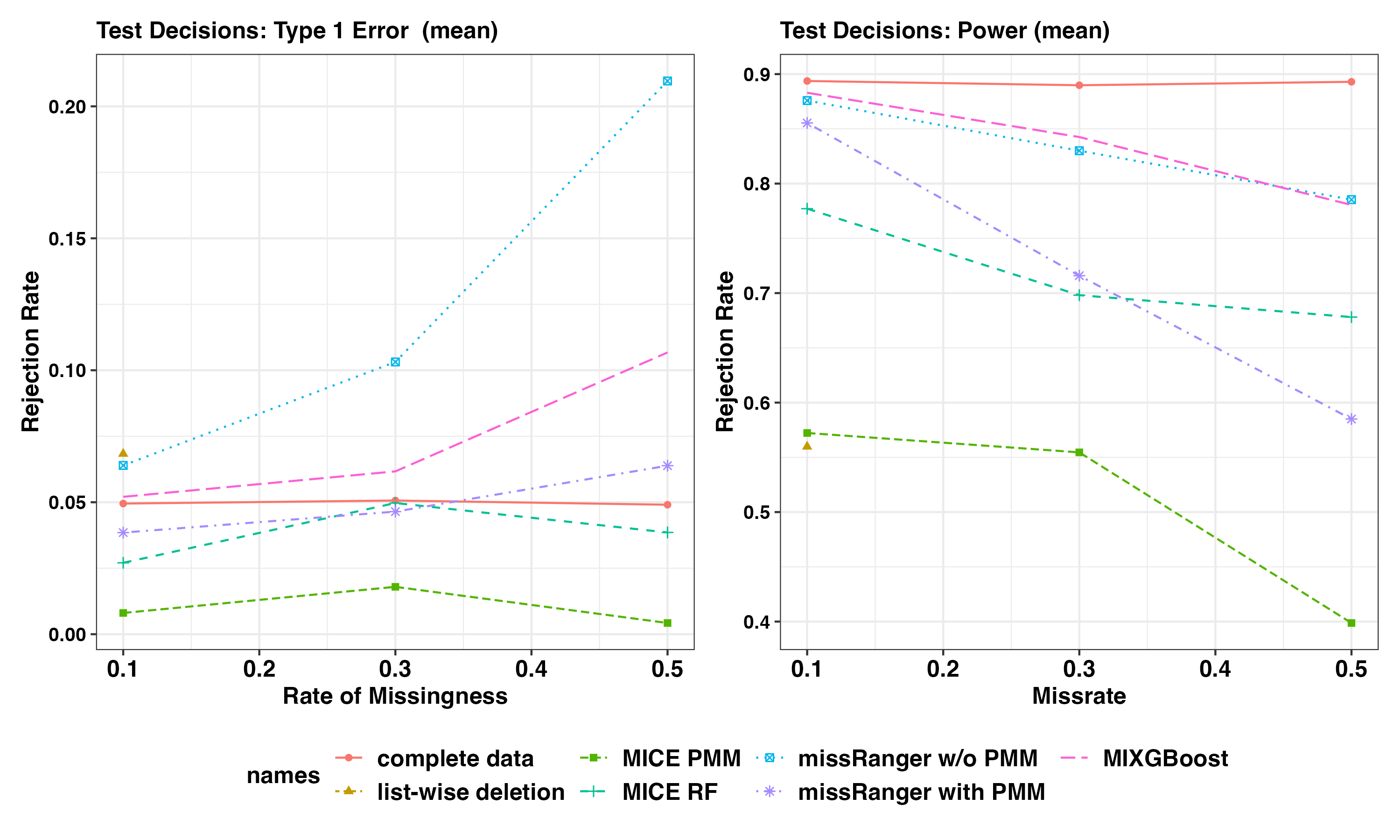}
 \end{subfigure}
     \begin{subfigure}{\textwidth}
        \centering
     \subcaption{Median Rejection Rates} \label{fig:devTDMedian}
 \includegraphics[width=0.85\textwidth]{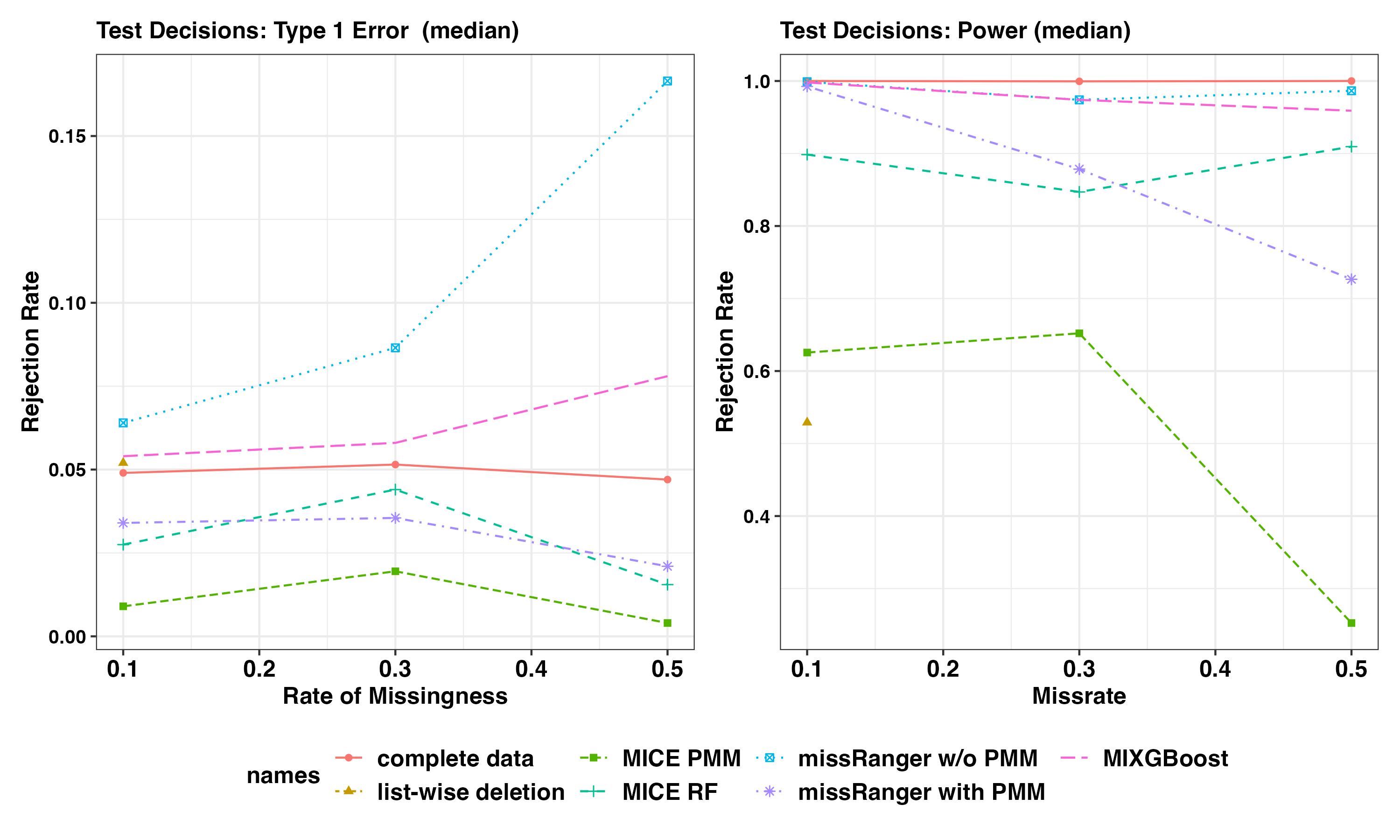}
  \end{subfigure}
  \caption*{\footnotesize{\textit{Note:} The Figure shows the rejection rates of the true coefficient, with higher values indicating a higher rejection rate. On the left for coefficients not equal to zero (as fixed in Equation \ref{eq1}), and on the right for coefficients equal to zero. "Complete" refers to estimation before any data are amputated, and "Deleted" refers to regression on listwise deleted datasets. The others refer to the imputation methods explained in the text.}}
\end{figure}

\clearpage \newpage
\subsection{IPM metric and running time results}

\begin{figure}[ht!]
    \centering
     \caption{IPM metric and running time results}
    \includegraphics[width=0.9\textwidth]{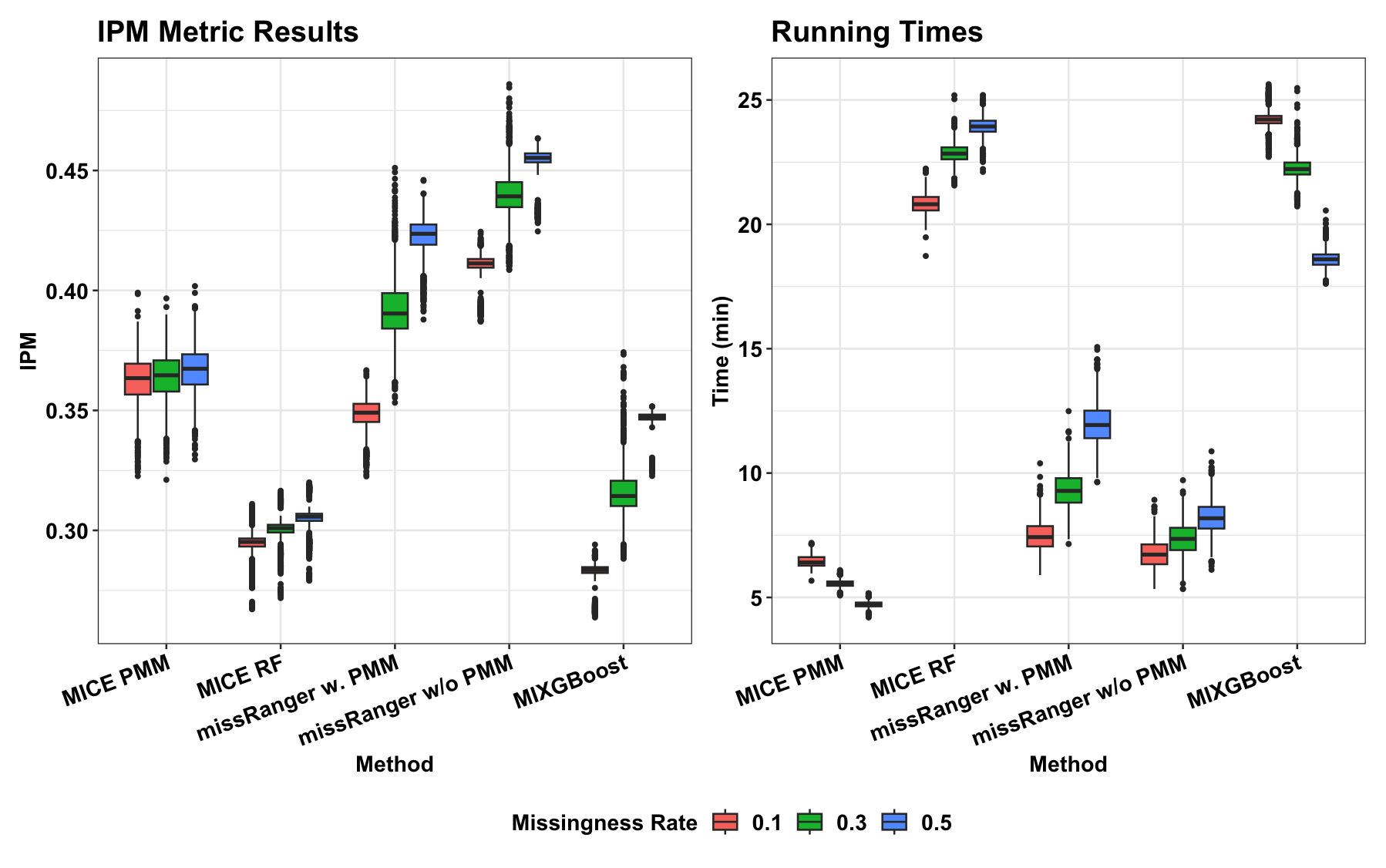}
    \label{fig:ipm}
\end{figure} 

\end{document}